\documentclass[11pt]{article}

\usepackage{longtable}
\usepackage{makeidx}
\usepackage{fullpage}
\usepackage{amsmath}

\DeclareMathOperator{\csch}{csch}
\usepackage{amssymb}
\usepackage{setspace}
\usepackage{bbm}
\usepackage{dsfont}
\usepackage{graphics}
\usepackage{epsfig}
\usepackage{mathtools}
\usepackage[font=footnotesize,labelfont=bf,justification=centerlast,width=.94\textwidth]{caption}
 \usepackage{multirow}
\usepackage{array,booktabs}
\usepackage{cite}
\usepackage{color}
\usepackage{float}
\usepackage[normalem]{ulem}

\usepackage{hyperref}

\hypersetup{
    bookmarks=true,%
    colorlinks,%
    citecolor=blue,%
    filecolor=black,%
    linkcolor=black,%
    urlcolor=blue
}

\newcommand{\be}{\begin{equation}}
\newcommand{\ee}{\end{equation}}
\newcommand{\bes}{\begin{equation*}}
\newcommand{\ees}{\end{equation*}}
\newcommand{\bea}{\begin{eqnarray}}
\newcommand{\eea}{\end{eqnarray}}
\newcommand{\beas}{\begin{eqnarray*}}
\newcommand{\eeas}{\end{eqnarray*}}

\newcommand{\bmat}{\begin{bmatrix}}
\newcommand{\emat}{\end{bmatrix}}

\newcommand{\RR}{\mathbb{R}}

\def\le{\left}
\def\ri{\right}

\def\le{\left}
\def\ri{\right}

\newcommand{\nn}{\nonumber}
\newcommand\fft[2]{\frac{#1}{#2}}
\newcommand\ft[2]{{\textstyle\frac{#1}{#2}}}

\newcommand{\tin}{\text{in}}
\newcommand{\tout}{\text{out}}

\begin{document}
\numberwithin{equation}{section}
{
\begin{titlepage}
\begin{center}

\hfill \\
\hfill \\
\vskip 0.75in

{\Large \bf Tweaking one-loop determinants in AdS$_3$}\\

\vskip 0.4in

{\large Alejandra Castro${}^{a}$, Cynthia Keeler${}^{b}$, and Phillip Szepietowski${}^{c}$}\\

\vskip 0.3in

${}^{a}${\it Institute for Theoretical Physics Amsterdam and Delta Institute for Theoretical Physics, University of Amsterdam, Science Park 904, 1098 XH Amsterdam, The Netherlands} \vskip .5mm
${}^{b}${\it Niels Bohr International Academy, Niels Bohr Institute, University of Copenhagen, Blegdamsvej 17, DK-2100 Copenhagen, Denmark} \vskip .5mm
${}^{c}${\it Institute for Theoretical Physics and Center for Extreme Matter and Emergent Phenomena,
	Utrecht University, Princetonplein 5, 3584 CC Utrecht, the Netherlands}
	%
\vskip .5mm
\texttt{a.castro@uva.nl, keeler@nbi.ku.dk, p.g.szepietowski@uu.nl}

\end{center}

\vskip 0.35in

\begin{center} {\bf ABSTRACT } \end{center}

We revisit the subject of one-loop determinants in AdS$_3$ gravity via the quasinormal mode method. Our goal is to evaluate a one-loop determinant with chiral boundary conditions for the metric field; chirality is achieved by imposing  Dirichlet boundary conditions on certain components while others satisfy Neumann. Along the way, we give a generalization of the quasinormal mode method for stationary (non-static) thermal backgrounds, and propose a treatment for Neumann boundary conditions in this framework. We evaluate the graviton one-loop determinant on the Euclidean BTZ background with parity-violating boundary conditions (CSS), and find excellent agreement with the dual warped CFT. We also discuss a more general falloff in AdS$_3$ that is related to two dimensional quantum gravity in lightcone gauge. The behavior of the ghost fields under both sets of boundary conditions is novel and we discuss potential interpretations. 
\vfill
\noindent \today

\end{titlepage}
}

\newpage

\tableofcontents

\section{Introduction}

One-loop corrections in holography provide a new window into the nature of quantum gravity. In AdS$_3$ the one-loop determinant of the graviton very elegantly establishes the anticipated results of Brown \& Henneaux \cite{Brown:1986nw}: finite (non-zero) energy excitations with Dirichlet boundary conditions fall into representations of the two dimensional conformal group.   This result was first argued in \cite{Maloney:2007ud,Yin2008}, and shown directly  in \cite{Giombi:2008vd} via heat  kernel methods. Since then, this subject in AdS$_3$ has been explored further, where the emphasis has been on either the inclusion of additional fields \cite{David:2009xg,DattaDavid2012} or modifications of the gravitational theory \cite{GaberdielGrumillerVassilevich2010,CastroLashkariMaloney2011,CastroLashkariMaloney2011a,BarnichGonzalezMaloneyEtAl2015}.

Our aim is to expand this discussion of one-loop determinants in AdS$_3$ gravity beyond the \emph{standard} Dirichlet boundary conditions. In particular, we will focus on a set of \emph{chiral} boundary conditions for the graviton: different components will satisfy either Dirichlet or Neumann boundary conditions such that the boundary theory has a fixed chirality.  Our motivation to carry out this computation is twofold. First, from a technical point of view we would like to present a concrete implementation of the evaluation of determinants with non-Dirichlet boundary conditions. Second, there is mounting evidence of interesting holographic interpretations of chiral boundary conditions for the metric in three dimensions; our one-loop corrections will provide a non-trivial holographic insight into these novel setups. 

We evaluate the one-loop contribution to the Euclidean path integral expanded as
\begin{align}
Z_{\rm grav}&= \int {\cal D} g\, e^{-{1\over \hbar} S[g]}\cr
&= \sum_{g_{\star}} \exp\le(-{1\over \hbar} S^{(0)}[g_{\star}]+ S^{(1)}[g_{\star}] +\hbar S^{(2)}[g_{\star}] + \cdots\ri)~.
\end{align}
Here $g$ should be viewed as a collection of fields including both metric and matter fields, and $S[g]$ is the corresponding Euclidean action for these fields. $g_{\star}$ corresponds to a classical saddle point around which we carry out a perturbative expansion in $\hbar$. $S^{(0)}$ corresponds to a tree level contribution, $S^{(1)}$ is the one-loop contribution and so forth. We will focus on $S^{(1)}$ exclusively. This contribution is controlled by suitable determinants of Laplacian operators (including any Fadeev-Popov determinants arising from gauge fixing); schematically we would write
\be
Z^{(1)}\equiv e^{S^{(1)}[g_\star]} = \det \le(\nabla^2_{g_\star} + m^2\ri)^{\pm}~,
\ee
where the $\pm$ refers to whether the determinant appears in the numerator (for fermions) or denominator (for bosons). As we mentioned above, there is an extensive literature on evaluating these determinants with Dirichlet boundary conditions, with one of the most canonical methods being the heat kernel.%
\footnote{Heat kernels very naturally have built-in Dirichlet boundary conditions: a basis of  normalizable eigenfunctions is used to describe a complete set of modes.} %
The heat kernel technique can be extended to include Neumann boundary conditions, as recently done in \cite{Giombi2013,GiombiKlebanovTseytlin2014} for higher dimensional AdS$_d$. The problem we face here is a mixture of both Dirichlet and Neumann, and while there might be a systematic way to adjust the heat kernel method to this setup, we will take a different route. 

The different route we will pursue is to \emph{tweak} the quasinormal mode method proposed by Denef-Hartnoll-Sachdev (DHS) \cite{DenefHartnollSachdev2010}. As we will review in section \ref{sec:2}, the original proposal of DHS is  based on analyticity: this leads to a concise expression for the functional determinant in a thermal geometry in terms of a product over quasinormal frequencies. The appearance of quasinormal frequencies in the product is directly tied to requiring Dirichlet boundary conditions for each component of the field in question. This feature allows us to tweak the DHS method to our agenda: by imposing instead Neumann boundary conditions on certain components of the graviton we will modify the spectrum of frequencies that enter in the functional determinant. This procedure will give us the control to treat each component of the metric individually as we implement the boundary conditions we are interested in.

There are two types of chiral boundary conditions we will study: CSS  \cite{CompereSongStrominger2013a,Troessaert2013b}, and $sl(2,\RR)$ Kac-Moody boundary conditions \cite{Avery2014,Apolo2014}. Both of these boundary conditions are characterised by allowing one piece of the boundary metric to fluctuate, while other components are fixed. In a nutshell the main features of these setups are:
\begin{description}
	\item[CSS:] These boundary conditions for AdS$_3$ are chosen such that the global symmetries inside the resulting asymptotic symmetry group become $ SL(2,\mathbb{R})_L\times U(1)_R$ instead of $ SL(2,\mathbb{R})_L\times SL(2,\mathbb{R})_R$. This smaller group of symmetries is motivated by the near horizon symmetries of extremal black holes. Implementating this condition leads to parity-violating boundary conditions, and as a result the $sl(2,\mathbb{R})_L\times sl(2,\mathbb{R})_R $ isometries of AdS$_3$ are only enhanced to a \emph{left-moving} Virasoro-$U(1)$-Kac-Moody algebra, with central charge $c$ and level $k$. A field theory with such a geometrical Virasoro-Kac-Moody structure is known as a Warped Conformal Field Theory (WCFT) \cite{HofmanStrominger2011,DetournayHartmanHofman2012}. 
	\item[$sl(2,\RR)$ KM:] This is a generalization of CSS that allows for more structure in the leading metric components while still being parity-violating. As a result the  $sl(2,\mathbb{R})_L\times sl(2,\mathbb{R})_R $ isometries of AdS$_3$ are enhanced to a \emph{left-moving} Virasoro plus an $sl(2,\RR)$ Kac-Moody algebra at level $k=c/6$. Unlike in CSS, one can improve the stress tensor such that we have zero central charge, and the Brown-York stress tensor vanishes. Thus this setup in AdS$_3$ is dual to  a two dimensional quantum gravity in lightcone gauge,  as elegantly argued in \cite{Apolo2014}, and not a conformal theory.
\end{description}
Since these boundary conditions are chiral (left-moving) in nature, to highlight their features we will need to implement the DHS method for stationary (not static) thermal backgrounds, i.e. for the Euclidean continuation of the rotating BTZ black hole.\footnote{The existing literature on using the DHS method is limited to static thermal backgrounds. The addition of angular momentum is not dramatic, but worth showing explicitly.}  As we evaluate the determinants in section \ref{sec:mixed}  and interpret them in section \ref{sec:holography}, the addition of  rotation will make evident that our derivations are unambiguously compatible with the dual description.

One of the most unexpected features in our derivations is the role of the ghost determinant contained in the graviton one-loop effective action. The role of the ghost fields is to remove states with zero energy from the path integral, i.e. to remove gauge redundancies. For Dirichlet boundary conditions one can see this explicitly after implementing the DHS prescription, and it is also in complete agreement with the heat kernel method. However, for the chiral boundary conditions we will use, the treatment of ghosts is more subtle: there are a priori two different conditions one can impose on eigenfunctions of the ghost, which dramatically change the resulting determinant for the graviton and its holographic interpretation. We will discuss these differences from the gravitational perspective (see section \ref{sec:mdngraviton}), and show how they affect the holographic interpretation in section \ref{sec:sl2hol}.

Our emphasis throughout will be on meromophic properties (the pole structure) of the one-loop determinant in AdS$_3$. There is in addition an entire function (a function that is holomorphic and has neither poles nor zeros) which we will not evaluate. Its purpose is to account for zero modes and contribute to the renormalization of various couplings. We will highlight in the main text when and where we are neglecting this piece and explore its role further in the discussion section. 

The outline of this paper is as follows. In section \ref{sec:2} we revisit the DHS method with Dirichlet boundary conditions for rotating BTZ, explicitly implementing the method on a stationary  background. In section \ref{sec:mixed} we consider chiral boundary conditions for the graviton, and evaluate the one-loop determinant on stationary backgrounds. The holographic interpretation of these determinants is discussed in  section \ref{sec:holography}. We close with a brief discussion in section \ref{sec:disc}. Appendix  \ref{app:btz} contains our conventions for the BTZ background, in appendix \ref{app:spin2}  we give a detailed study of the spin-2 fluctuations, and in appendix  \ref{app:spin1} we describe the ghost  spectrum.


\section{ Quasinormal mode method: Rotation }\label{sec:2}

In this section we show how to implement the DHS prescription in spacetimes which are rotating, i.e. they are stationary but not necessarily static. We begin with a generic discussion and then provide concrete examples for one-loop determinants of massive and massless fields on the rotating BTZ black hole background.

The main proposition of DHS \cite{DenefHartnollSachdev2010} is a formula for functional determinants in a thermal geometry, written as a product over quasinormal frequencies.  Their arguments rely on the assumption of meromorphicity of the determinant in the mass parameter.  For example, consider the one-loop determinant on a thermal background, such as a Euclidean AdS black hole.%
\footnote{For sake of simplicity we will limit the discussion to black hole backgrounds in AdS. The method of DHS applies more generally and the discussion in this section should be valid for those cases. We are also setting the AdS radius to one throughout.}
For a complex scalar field we have
\be
Z^{(1)}(\Delta)=\int {\cal D}\varphi \exp \left(-\int d^{d+1}x\,\sqrt{g}\,\varphi^*(-\nabla^2+m^2)\varphi \right)=\det\left(-\nabla^2+m^2\right)^{-1}~,\quad \Delta(\Delta-d)= m^2~.
\ee 
If $Z^{(1)}(\Delta)$ analytically continued to the complex $\Delta$ plane is a meromorphic function, then it can be characterized by the locations and degeneracies of its poles and zeros, as well as its behavior at infinity.  For a scalar, $Z^{(1)} \propto \det^{-1}$, so there are no zeros.  Poles occur when $\det =0$, which happens whenever $\Delta$ is tuned such that a zero mode of $\varphi$ exists.

In Euclidean space, zero modes%
\footnote{Note that these Euclidean zero modes generically occur at non-physical values of $\Delta$; in other words they do not correspond to \emph{actual} zero modes in the evaluation of the one-loop determinant.} %
 are solutions of the Klein-Gordon equation which are smooth, obey the given asymptotic boundary conditions, and are single-valued in the Euclidean time direction.  We denote these solutions by $\varphi_{\star,n}$ where $n$ labels the mode number in the Euclidean time direction, and $\star$ labels all other quantum numbers characterizing the solution.  A given $\varphi_{\star,n}$ will solve the Klein-Gordon equation only when $\Delta$ is tuned to a particular value dependent on these quantum  numbers; we call this value $\Delta_{\star,n}$.  Thus from the Euclidean perspective, poles in $Z^{(1)}$ occur at all $\Delta=\Delta_{\star,n}$; if multiple sets of quantum numbers give Klein-Gordon solutions with the same value of $\Delta_{\star,n}$ then the pole is accordingly of higher multiplicity.

The key insight of DHS is to relate the Euclidean zero modes $\varphi_{\star,n}$ to Lorentzian quasinormal modes via Wick rotation.  The Euclidean thermal spacetime Wick-rotates to a black hole spacetime.  From this Lorentzian perspective, (anti)quasinormal modes are solutions to the Klein-Gordon equation satisfying (out)ingoing boundary conditions at the black hole horizon, as well as normalizable asymptotic boundary conditions. These modes can be characterized by their (anti)quasinormal frequencies $\omega_\star(\Delta)$, where $\star$ represents the spatial quantum numbers.  Importantly these frequencies depend on $\Delta$, and we find
\be\label{omegarelation}
\omega_\star(\Delta_{\star,n})=\omega_n=2\pi i n T,
\ee
when we tune $\Delta=\Delta_{\star,n}$.   At these specific values, each Lorentzian quasinormal mode $\varphi_{\star,\omega}$ Wick-rotates into the Euclidean zero mode $\varphi_{\star,n}$, with $n\geq 0$.  The second equality here holds only for static black holes, where the thermal frequency $\omega_n$ relates directly to the Euclidean mode number $n$.  
In this case, the condition of smoothness near the vanishing of the thermal cycle in the Euclidean space Wick-rotates to the ingoing condition at the horizon of the Lorentzian space. 

For $n<0$, the Euclidean modes instead match onto outgoing quasinormal modes (or antiquasinormal modes). For the ``constant'' modes with $n=0$, one can work with either in or outgoing quasinormal modes.  

Consequently, if we know all of the quasinormal and antiquasinormal frequencies as a function of $\Delta$, we know the poles in $Z^{(1)}(\Delta)$ will be located where $\Delta$ is tuned such that $\omega_\star(\Delta)=\omega_n$.  We can now write the determinant for the complex scalar as
\be\label{eq:zdhs}
Z^{(1)}(\Delta)= e^{{\rm Pol}(\Delta)}\prod_{n,\star} \left(\omega_n-\omega_\star(\Delta)\right)^{-1}~.
\ee
Here the product is over all quantum numbers that control the (anti)quasinormal frequencies, denoted succinctly by ``$\star$.'' We have also included an entire function (that is, a function that is holomorphic and has neither poles nor zeros), via $e^{{\rm Pol}(\Delta)}$ where ${\rm Pol}(\Delta)$ is a polynomial with only positive powers in $\Delta$.  We can determine this polynomial separately.\footnote{In \cite{DenefHartnollSachdev2010} ${\rm Pol}(\Delta)$ is determined by matching the $\Delta\to\infty$ behavior. We will not focus on this contribution in the following, but we will comment on it in our discussion. } 

In this section we want to implement \eqref{eq:zdhs} for stationary backgrounds, and in particular rotating black holes. The minor tweak we need to implement is to revisit the Euclidean regularity condition, which affects the thermal frequencies $\omega_n$; the second equality in \eqref{omegarelation} will change. For a suitable radial coordinate $R$ and Euclidean time coordinate $T_E$, the metric near the horizon will take the form
\be
ds^2 \approx dR^2 +R^2 dT_E^2 + ds^2_{\perp}~, \qquad T_E\sim T_E +2\pi ~,
\ee 
in a similar fashion as for the static solution.  However, for a rotating background at temperature $T$ and angular velocity $\Omega$, the Wick rotation to Lorentzian signature is generically of the form $T_E= 2\pi T (i t +  \Omega \phi) $, where $\phi$ is the axis of rotation of the black hole. This implies that regularity of the fields at $R=0$ will impose a condition on quantum numbers conjugate to both $\partial_t$  and $\partial_\phi$. In the following we will work out explicit examples to illustrate this modification.

\subsection{Example: real scalar field on BTZ black hole}\label{sec:Dscalar}

As a warmup, in this subsection we evaluate the one-loop determinant for a massive real scalar field on the rotating BTZ black hole with Dirichlet boundary conditions. This should be contrasted  with the static case done in \cite{DenefHartnollSachdev2010}; see \cite{Giombi:2008vd,David:2009xg} for a derivation using heat kernel methods. 

To start, we impose Dirichlet asymptotic boundary conditions on scalar field solutions
\begin{equation}
\varphi(r,t,\phi) \sim r^{-\Delta} e^{-i\omega t + i k \phi}
\end{equation}
for large values of $r.$ Here we have written the Fourier mode with frequency $\omega$ and wave number $k$, as appropriate for the coordinate system \eqref{eq:coordbl}.%
\footnote{All relevant details about the background metric are listed in appendix \ref{app:btz}.} 
In addition, periodicity in the $\phi$ coordinate restricts the wave number $k$ to take values over all of the integers.

Now, let us consider the behavior of the Lorentzian solution for the scalar field near the horizon, $r\sim r_+$:
\be\label{eq:Lorsol}
\varphi(r,t,\phi)\sim (r-r_+)^{\pm i {k_T \over 2}} e^{-i\omega t + i k \phi}~,\qquad k_T = {\omega \,r_+ -k\, r_-\over r_+^2-r_-^2}~.
\ee
The dependence on $k_T$ is set by the equations of motion, where $k_T$ is defined as the frequency conjugate to the coordinate\footnote{This is not to be confused with the temperature $T$ mentioned previously. We hope that context will be enough to distinguish between the two meanings.} $T$ as specified in \eqref{eq:klkr}. For general values of  $\omega$ and $k$, solutions satisfying the boundary conditions at $r \rightarrow \infty$ will have both of the $(r-r_+)^{\pm i {k_T \over 2}}$ behaviors near the horizon. Solutions which satisfy only one of the behaviors in (\ref{eq:Lorsol}) occur only at specific quantized values of the frequency $\omega;$ depending on the sign of $k_T$ in (\ref{eq:Lorsol}) these are the quasinormal and antiquasinormal frequencies.

Wick-rotating to $T_E=iT$ and changing to the regular Euclidean coordinates \eqref{eq:regcoord} near $\xi=0$, the solutions in (\ref{eq:Lorsol}) become
\be
\varphi(\xi,T_E,\Phi)\sim \xi^{\pm i k_T} e^{- k_T T_E}e^{-i k_\Phi \Phi}~.
\ee
Regularity of these solutions requires that $k_T=in$, where $n$ is any integer.  Additionally if $n\geq 0$, we must have only the $\xi^{- i k_T}$ behavior; if $n\leq 0$ we instead have $\xi^{+ i k_T }$.%
\footnote{For $n=0$ we may choose to treat it as either $\pm$, that is either antiquasinormal or quasinormal; the important condition for quasinormal modeness here is that we do not allow the $\log$ behavior that would arise for general $\omega,k$.  Note that as for non-rotating (static) case, the quasinormal mode spectrum here satisfies $\prod (\omega_0-\omega_{\star,\rm in})=\prod \sqrt{ (\omega_0-\omega_{\star,\rm in})  (\omega_0-\omega_{\star,\rm out})}$, so we can indeed choose to treat $n=0$ modes together with either the quasinormal or antiquasinormal frequencies. We will treat $n=0$ with whichever case is most convenient in the following (usually with the quasinormal modes). We will also refer to both quasinormal and antiquasinormal modes with just the word quasinormal, specifying instead either the sign of $n$ or the ingoing/outgoing nature of the mode in question.}
Choosing only one of these signs in the Lorentzian solution \eqref{eq:Lorsol} amounts to choosing either ingoing (for $n\geq 0$) or outgoing (for $n\leq 0$) conditions at the horizon; thus, the solutions we are interested in should be either quasinormal or antiquasinormal modes.

In addition, the requirement to have $k_T=in$ forces the (anti)quasinormal frequency $\omega$ to take the specific value $\omega_n$:
\be\label{eq:omegarot}
-i k_T =n  \quad \Rightarrow \quad {\omega _n\over 2\pi} = 2i {T_LT_R\over T_L+T_R} n + {T_R-T_L\over T_L +T_R} {k\over 2\pi}~,
\ee
where
\be
T_L={1\over 2\pi} (r_+-r_-) ~,\qquad T_R={1\over 2\pi}(r_+ + r_-)~.
\ee

Next, the quasinormal frequencies of a real scalar field on the background of a rotating BTZ black hole are \cite{Cardoso2001,Birmingham2002,Berti2009}
\begin{center}
	\begin{tabular}{|c|c|}
		\hline
		ingoing & outgoing\\
		\hline
		$\begin{matrix}\, \omega_\star = -k - 2\pi i T_R (2p + \Delta )  \,\,\, \\
		\,  \omega_\star = k - 2\pi i T_L(2p + \Delta )  \,\,\,\end{matrix}$&  
		$\begin{matrix}\, \omega_\star = -k + 2\pi i T_R (2p + \Delta )   \,\,\, \\
		\,\omega_\star = k + 2\pi i T_L(2p + \Delta )   \,\,\,\end{matrix}$\\ 
		\hline
	\end{tabular}
\end{center}
The range of $k$ is all integers, and $p$ is a nonnegative integer. Implementing \eqref{eq:zdhs}, the one-loop determinant of a scalar field  on the background of rotating BTZ becomes
\bea\label{eq:prodrot}
\le({e^{\rm Pol(\Delta)} \over Z^{(1)}}\ri)^2 &=& \prod_{n>0,p\geq0,k} \le (\omega_n+ k +2\pi i T_R (2p + \Delta )\ri)\le (\omega_n- k + 2\pi i T_L(2p + \Delta ) \ri)\cr
&& \prod_{n<0,p\geq0,k} \le (\omega_n+k - 2\pi i T_R (2p + \Delta )\ri)\le (\omega_n- k - 2\pi i T_L(2p + \Delta )\ri)\cr
&&\prod_{p\geq0,k} \le (\omega_0+k + 2\pi i T_R (2p + \Delta )\ri)\le (\omega_0-k + 2\pi i T_L(2p + \Delta )\ri)~,
\eea
where $\omega_n$ is given by \eqref{eq:omegarot}. Note that we want the determinant for a real scalar, hence the square on the left hand side of \eqref{eq:prodrot}. The first line in \eqref{eq:prodrot} corresponds to the ingoing modes hitting thermal frequencies with $n>0$, the second line are the outgoing modes and thermal frequencies with $n<0$, and the last line corresponds to the zero modes with $n=0$. After plugging in $\omega_n$ and a bit of algebra, we have
\bea\label{eq:prodrot1}
\le({e^{\rm Pol(\Delta)} \over Z^{(1)}}\ri)^2 &=& \prod_{n>0,p\geq0,k} \le( \le(p+{\Delta\over 2} + n {T_L\over T_L+T_R}\ri)^2  + \le({k\over 2\pi( T_L+T_R)}\ri)^2\ri) \cr &&\prod_{n>0,p\geq0,k}\le( \le(p+{\Delta\over 2} + n {T_R\over T_L+T_R}\ri)^2  + \le({k\over 2\pi(T_L+T_R)}\ri)^2\ri)\cr
&&\prod_{p\geq0,k} \le( \le(p+{\Delta\over 2} \ri)^2  + \le({k\over 2\pi(T_L+T_R)}\ri)^2\ri)~.
\eea
Next, we regulate the product over $k$ by using the formula
\be\label{eq:kProd}
\prod_{k>0}\le(1+{x^2\over k^2}\ri)={\sinh{\pi x}\over \pi x}={e^{\pi x}\over \pi x} (1-e^{-2\pi x})~, 
\ee
which, up to a redefinition of $\rm Pol(\Delta)$, turns \eqref{eq:prodrot1} into%
\footnote{The $k=0$ terms in (\ref{eq:prodrot1}) conveniently cancel the various terms that appear due to the denominator of \eqref{eq:kProd} which are not entire functions of $\Delta.$}
\bea\label{eq:prodrot2}
{e^{\rm Pol(\Delta)} \over Z^{(1)}} &=& \prod_{n>0,p\geq0} \le( 1- q^{n+p} \bar q^p (q \bar q)^{\Delta/2}\ri) \cr &&\prod_{n>0,p\geq0}\le(1- \bar q^{n+p}  q^p (q \bar q)^{\Delta/2}\ri) \prod_{p\geq0}\le(1-  (q \bar q)^{p+\Delta/2}\ri)~,
\eea
where we defined\footnote{In terms of the complex structure $\tau$, we would have $q=e^{2\pi i\tau}$ and  $\bar q=e^{-2\pi i\bar\tau}$, where $\tau=2\pi iT_L$ and $\bar \tau= - 2\pi i T_R$. Note that in Euclidean signature $(T_L)^* =  T_R $ since $r_-$ is purely imaginary.} 
\be
q\equiv e^{-2\pi (2\pi T_L)}~,\qquad \bar q\equiv e^{-2\pi (2\pi T_R)}~.
\ee
Rewriting \eqref{eq:prodrot2}, the answer for the one-loop determinant of a real scalar is
\be\label{eq:GMY}
Z^{(1)}= e^{\rm Pol(\Delta)}\prod_{\ell,\ell'=0}^\infty{1\over (1-q^{\ell +\Delta/2}\bar q^{\ell' +\Delta/2})}~,
\ee
in complete agreement with \cite{Giombi:2008vd,David:2009xg}, and with \cite{DenefHartnollSachdev2010} for the static solution. Note that despite appearances, \eqref{eq:GMY} is equal to \eqref{eq:prodrot2}. 
One heuristic way to see this is as follows: the first product in \eqref{eq:prodrot2} corresponds to $\ell >\ell'$, the second product is $\ell < \ell'$ and the last product is $\ell=\ell'$. 

To fully specify the one-loop determinant one should also determine the $e^{\rm Pol(\Delta)}$ factor in (\ref{eq:GMY}). This term corresponds to a local renormalization of the classical action and can be computed independently in a suitable large-$\Delta$ limit, for example by using heat kernel techniques as described in \cite{DenefHartnollSachdev2010}. This result can then be matched to the large-$\Delta$ limit of expressions such as (\ref{eq:GMY}) to determine $\rm Pol(\Delta).$ In this paper we are specifically interested in the properties of the infinite products that occur in the one-loop determinant, such as that in (\ref{eq:GMY}). The location of the poles that occur in these products are independent of $\rm Pol(\Delta)$ and so we will often drop the $e^{\rm Pol(\Delta)}$ factor completely. In the remainder of this paper, expressions for one-loop determinants should be understood to correspond to the determinant modulo these local renormalization terms. We will only comment on $\rm Pol(\Delta)$ in cases where determining it may be subtle.


\subsection{Example: spin-2 fields on BTZ black hole}\label{subsec:gravBTZ}

As a second example we would like to illustrate how to evaluate the one-loop determinant for spin-2 fields, both massive and massless. References \cite{DattaDavid2012,ZhangZhang2012} discuss this evaluation via the quasinormal mode method for the static case, and we follow closely their analysis of the Fronsdal equations. We add the evaluation  of  the determinants for rotating backgrounds, and an improved discussion on how to identify the set of frequencies $\omega_\star$ that control the poles of $Z(\Delta)$.

Following \cite{DattaDavid2012}, a massive spin-2 excitation $h_{\mu\nu}$ in AdS$_3$ satisfies the first order equation
\begin{eqnarray}\label{eq:firstorderspin2}
\epsilon_\mu{}^{\alpha\beta}\nabla_{\alpha}h_{\beta\nu} = - m_2 \,h_{\mu\nu}~,
\end{eqnarray}
where the sign of $m_2$ controls the helicity of the field. Using the equations for both helicities, it follows that such a field satisfies the more familiar Fronsdal equations: 
\begin{eqnarray}\label{eq:firstorderspin2both}
\nabla^\mu h_{\mu\nu} &=& 0~,\nn\\
h^\mu{}_\mu &=& 0 ~, \nn\\
\nabla^2 h_{\mu\nu} &=&  (m_2^2 - 3)h_{\mu\nu}~.
\end{eqnarray}
For $m_2 = \pm 1,$ these are the equations of motion for linearized graviton fluctuations. The physical graviton has two degrees of freedom corresponding to positive and negative states, one for each sign of $m_2$. Setting $m_2=\pm1$ we identify $\delta g_{\mu\nu} = h_{\mu\nu},$ where $\delta g_{\mu\nu}$ is restricted to be a transverse and traceless metric fluctuation. 

The determinant we will evaluate is 
\be\label{eq:zm2}
Z^{(1)}_{s=2}(\Delta_2) = \le({\rm det}_{\rm STT}(-\nabla^2 + m_2^2-3)\ri)^{-1/2}~, \qquad \Delta_{2} \equiv |m_2| + 1~.
\ee
We emphasise that  $\nabla^2$ in \eqref{eq:zm2} is acting on a symmetric, traceless and transverse tensor. In this section we evaluate the determinant for  standard (Dirichlet) boundary conditions: the leading divergence%
\footnote{The leading divergence here refers to the leading behavior at physical values of $\Delta$.  Schematically, this means we allow $r^\Delta$ behavior but not $r^{d-\Delta}$.  Since we are formally studying the determinant throughout the $\Delta$ complex plane, this condition differs slightly from normalizability.  Instead it is the natural analytic continuation of normalizability.  We will not encounter this subtlety here as we are in odd dimensional AdS; consequently we will use ``normalizable'' to refer to the analytic continuation. For more details, see \cite{KeelerNg2014,KeelerLisbaoNg2016}.}
 of the zero modes at the boundary is required to vanish, which is the usual condition for quasinormal modes in AdS.  We provide a detailed derivation of the spin-$2$  quasinormal modes, as well as the mapping to regular Euclidean solutions, in appendix \ref{app:spin2}.  Here, we only quote the results for the quasinormal mode spectra; for spin-2 these are in Table \ref{table:BHQNMspin2}.

\noindent\begin{minipage}{\textwidth}
	\begin{center}
		\begin{tabular}{|c|c|c|}
			\hline
			& ingoing & outgoing\\
			\hline
			$m_2 > 0$	& $\begin{matrix}\, 2ik_R = 2p + \Delta_{2} + 2  \,\,\, \\
			\,2ik_L =  2p + \Delta_{2} - 2  \,\,\,\end{matrix}$&  $\begin{matrix}\, 2ik_R = -(2p + \Delta_{2} + 2)  \,\,\, \\
			\,2ik_L = -(2p + \Delta_{2} - 2)  \,\,\,\end{matrix}$\\ 
			\hline
			$m_2 < 0$ & $\begin{matrix}\, 2ik_R = 2p + \Delta_{2} - 2  \,\,\, \\
			\,2ik_L =  2p + \Delta_{2} + 2  \,\,\, \end{matrix}$ & $\begin{matrix}\, 2ik_R = -(2p + \Delta_{2} - 2)  \,\,\, \\
			\,2ik_L =  -(2p + \Delta_{2} + 2)  \,\,\,\end{matrix}$\\
			\hline
		\end{tabular}\captionsetup{justification=raggedright}\captionof{table}{Spin-$2$ quasinormal mode spectrum $\omega_\star$ after imposing standard Dirichlet boundary conditions. When $\Delta_{2}=2$ these correspond to the symmetric, transverse, traceless graviton spectrum. Each condition on $k_R$ or $k_L$ labels a distinct eigenmode and the range of $p$ is over all non-negative integers.}\label{table:BHQNMspin2}
	\end{center}
	\vspace{.2em}
\end{minipage}
Note that we are parameterizing the quasinormal frequencies in terms of the quantum numbers $(k_L,k_R)$ as defined in \eqref{eq:klkr}, which are conjugate to the coordinates $(x_L,x_R)$ in \eqref{DDsec3.2}. In the following we will also use $(k_T,k_\Phi)$ whose conjugate variables are $(T,\Phi)$ in \eqref{eq:defnTP}.

Next, we need to match the quasinormal frequencies to the thermal frequencies, i.e. $\omega_n=\omega_\star$. Additionally, some of the quasinormal modes with low $p$ and $n$ Wick-rotate to Euclidean modes that diverge at the tip of the Euclidean cigar, so they should be excluded.
The relations defining the good Euclidean solutions are enumerated in Appendix \ref{app:spin2Eucl}, and are reproduced in the following table: 

\noindent\begin{minipage}{\linewidth}
	\begin{center}
		\begin{tabular}{|c|c|}
			\hline
			$m_2>0$ & $m_2<0$ \\
			\hline
			$\begin{matrix}\, 2p + \Delta_2 + |n+2| + ik_\Phi(n,k) = 0  \,\,\, \\
			\,  2p + \Delta_2 + |n-2| - ik_\Phi(n,k) = 0   \,\,\,\end{matrix}$&  
			$\begin{matrix}\, 2p + \Delta_2 + |n-2| + ik_\Phi(n,k) = 0   \,\,\, \\
			\,2p + \Delta_2 + |n+2| - ik_\Phi(n,k) = 0 \,\,\,\end{matrix}$\\ 
			\hline
		\end{tabular}\captionsetup{justification=raggedright}\captionof{table}{Conditions satisfied by Euclidean solutions with standard quasinormal boundary conditions. Each solution satisfies one of the conditions listed. Here $k_\Phi(n,k)$ is given in equation (\ref{eq:kPhi}). In this table, $p$ runs over all non-negative integers, whereas $n$ and $k$ run over all integers.}\label{table:BHEuclCond}
	\end{center}
	\vspace{.2em}
\end{minipage}

As in the previous example, $n$ is defined by the regularity condition at the Euclidean origin, which fixes $-ik_T=n$. Each set of conditions corresponds to a union of the Wick-rotation of a set of ingoing and outgoing states. Ingoing modes correspond to $n>0$ and outgoing modes to $n<0,$ with $n=0$ being the zero mode.  
The frequency $k_\Phi$ is restricted by the periodicity of the field in the thermal and spatial directions, which is controlled by integers $n$ and $k$ respectively. The relation is
\begin{equation}\label{eq:kPhi}
k_\Phi(n,k) = \frac{T_R-T_L}{T_R+T_L} i  n - \frac{1}{T_R+T_L}\frac{k}{\pi}~.
\end{equation}

We can construct the determinant directly from this information. Consider first the $m_2>0$ states: the conditions from the top row in Table \ref{table:BHEuclCond} can be written as
\begin{equation}
2p + \Delta_{2} + |n + 2| - \frac{T_R-T_L}{T_R+T_L} n - \frac{1}{T_R+T_L}\frac{i k}{\pi} = 0~.
\end{equation}
Relabelling $n = \tilde n - 2$ and treating each sign separately, we have
\begin{align}\label{eq:LMspin2Poles}
2 p + \Delta_{2} + 2\frac{T_R-T_L}{T_R+T_R} + \frac{2T_L}{T_R+T_R}\tilde{n} - \frac{1}{T_R+T_L}\frac{i k}{\pi} = 0, &\qquad \tilde{n} > 0~, \nn \\
2 p + \Delta_{2} + 2\frac{T_R-T_L}{T_R+T_R} - \frac{2T_R}{T_R+T_R}\tilde{n} - \frac{1}{T_R+T_L}\frac{i k}{\pi} = 0, &\qquad \tilde{n} < 0~, \nn \\
2 p + \Delta_{2} + 2\frac{T_R-T_L}{T_R+T_R} - \frac{1}{T_R+T_L}\frac{i k}{\pi} = 0, &\qquad \tilde{n} = 0~.
\end{align}
Performing similar steps for the $m_2>0$ states on the bottom row in Table \ref{table:BHEuclCond}, using instead $n = \tilde{n} +2,$ we find
\begin{align}\label{eq:RMspin2Poles}
2 p + \Delta_{2} + 2\frac{T_R-T_L}{T_R+T_R} + \frac{2T_R}{T_R+T_R}\tilde{n} + \frac{1}{T_R+T_L}\frac{i k}{\pi} = 0, &\qquad \tilde{n} > 0~, \nn \\
2 p + \Delta_{2} + 2\frac{T_R-T_L}{T_R+T_R} - \frac{2T_L}{T_R+T_R}\tilde{n} + \frac{1}{T_R+T_L}\frac{i k}{\pi} = 0, &\qquad \tilde{n} < 0~, \nn \\
2 p + \Delta_{2} + 2\frac{T_R-T_L}{T_R+T_R} + \frac{1}{T_R+T_L}\frac{i k}{\pi} = 0, &\qquad \tilde{n} = 0~.
\end{align}

We can compare these conditions with those imposed on the zeros of the expression (\ref{eq:prodrot1}). If one makes the replacement 
\begin{equation}\label{eq:Deltashiftm>0}
\Delta \rightarrow \Delta_{2} + 2\frac{T_R-T_L}{T_R+T_L}~,
\end{equation}
in (\ref{eq:prodrot1}), and also replaces the $n$ in (\ref{eq:prodrot1}) with $n=|\tilde{n}|,$ one precisely reproduces the conditions in (\ref{eq:LMspin2Poles}) and (\ref{eq:RMspin2Poles}) from the zeros in (\ref{eq:prodrot1}). Therefore, we can determine the result for the spin-$2$ determinant from the real scalar case (\ref{eq:GMY}) by making the replacement (\ref{eq:Deltashiftm>0}), which gives
\begin{eqnarray}
Z^{(1)}_{m_2>0} &=&  \left(\det\nolimits_{\rm STT} (-\nabla^2 +m_2^2 - 3)_{m_2>0}\right)^{-1/2}\nn\\
&=&  \prod_{\ell,\ell'=0}^\infty{1\over (1-q^{\ell +h}\bar q^{\ell' + h + 2})}~,
\end{eqnarray}
where $h$ is the weight of the spin-2 field, given by
\begin{equation}
\Delta_2=2h +2~.
\end{equation}

It is now straightforward to also read off the contribution from the $m_2<0$ states. Since the only difference from the $m_2>0$ case is on the sign of $k_\Phi,$ the $m_2<0$ result will be the same but with the opposite shift 
\begin{equation}
\Delta \rightarrow \Delta_{2} - 2\frac{T_R-T_L}{T_R+T_L}~,
\end{equation}
which leads to 
\begin{equation}
Z^{(1)}_{m_2<0}= \prod_{\ell,\ell'=0}^\infty{1\over (1-q^{\ell +h+2}\bar q^{\ell' + h})}~.
\end{equation}

Putting it all together we arrive at the entire one-loop massive spin-$2$ determinant 
\begin{eqnarray}\label{eq:BHspin2det}
Z^{(1)}_{s=2} &=& Z^{(1)}_{m_2>0}\,Z^{(1)}_{m_2<0} \nn\\
&=& \prod_{\ell,\ell'=0}^\infty{1\over (1-q^{\ell +h+2}\bar q^{\ell' + h}) (1-q^{\ell +h}\bar q^{\ell' + h + 2})}~.
\end{eqnarray}
This agrees with the results in \cite{Giombi:2008vd,David:2009xg}, which were derived using heat kernel methods, and with \cite{DattaDavid2012} when the rotation is turned off. We will postpone the holographic interpretation of these determinants to section \ref{sec:holography}.

\subsubsection{Graviton determinant}\label{sec:gdd}

In this section we are interested in the standard Dirichlet boundary conditions for the graviton, which corresponds to allowing only fluctuations which fall off at least as fast as
\begin{eqnarray}\label{eq:bhbc}
\delta g_{\mu\nu} \sim \mathcal O(r^{0})~,
\end{eqnarray}
near the AdS boundary. Since there are extra gauge redundancies in the massless case, we need to include as well the well-known ghost determinant. Hence, the graviton one-loop determinant is \cite{GibbonsPerry1978,ChristensenDuff1980,Yasuda1984}
\begin{equation}\label{eq:grav-det}
Z^{(1)}_{\rm grav} = \left(\frac{\det_{\rm T}(-\nabla^2 + 2/L^2)}{\det_{\rm STT}(-\nabla^2 - 2/L^2)}\right)^{1/2}~,
\end{equation}
where the denominator is the determinant for symmetric, tranverse and traceless rank-2 tensors and the numerator is the determinant for transerve vector fields. These determinants correspond to fields with physical mass values $m_{2}^2 = 1 $ for the graviton and $m_{1}^2 = 4$ for the ghost; the corresponding  conformal dimensions are
\begin{equation}
\Delta_{2} = 2~, \qquad \Delta_{1} = 3~.
\end{equation}

Let us first evaluate the numerator in \eqref{eq:grav-det} for spin-$1$ fields with arbitrary $\Delta_1$ and standard boundary conditions. The quasinormal mode spectrum of a vector field in AdS$_3$ is derived in appendix \ref{app:spin1}; the resulting frequencies are listed in the table below.

\noindent\begin{minipage}{\linewidth}
	\begin{center}
		\begin{tabular}{|c|c|c|}
			\hline
			& ingoing & outgoing\\
			\hline
			$m_1 > 0$	& $\begin{matrix}\, 2ik_R = 2p + \Delta_{1} + 1  \,\,\, \\
			\,2ik_L =  2p + \Delta_{1} - 1  \,\,\,\end{matrix}$&  $\begin{matrix}\, 2ik_R = -(2p + \Delta_{1} + 1)  \,\,\, \\
			\,2ik_L = -(2p + \Delta_{1} - 1)  \,\,\,\end{matrix}$\\ 
			\hline
			$m_1 < 0$ & $\begin{matrix}\, 2ik_R = 2p + \Delta_{1} - 1  \,\,\, \\
			\,2ik_L =  2p + \Delta_{1} + 1  \,\,\, \end{matrix}$ & $\begin{matrix}\, 2ik_R = -(2p + \Delta_{1} - 1)  \,\,\, \\
			\,2ik_L =  -(2p + \Delta_{1} + 1)  \,\,\,\end{matrix}$\\
			\hline
		\end{tabular}
		\captionsetup{justification=raggedright}\captionof{table}{Spin-$1$ quasinormal mode spectrum after imposing standard Dirichlet boundary conditions. When $\Delta_{1}=3$ these correspond to the spectrum of transverse ghost modes which appear in the graviton one-loop determinant (\ref{eq:grav-det}). Each condition on $k_R$ or $k_L$ labels a distinct eigenmode and the range of $p$ is over all non-negative integers. }\label{table:BHQNMspin1}
	\end{center}
	\vspace{.2em}
\end{minipage}

\noindent We can derive the spin-$1$ contribution to the determinant similarly to the spin-$2$ case (\ref{eq:BHspin2det}). The general result for the determinant of a massive spin-1 field is
\begin{eqnarray}\label{eq:BHspin1det}
Z^{(1)}_{s=1} &=& Z^{(1)}_{m_1>0}\,Z^{(1)}_{m_1<0} \nn\\
&=& \prod_{\ell,\ell'=0}^\infty{1\over (1-q^{\ell +h+1}\bar q^{\ell' + h}) (1-q^{\ell +h}\bar q^{\ell' + h + 1})}~.
\end{eqnarray}
For a spin-$1$ field $\Delta_{1} = 2 h + 1$ and the contribution in (\ref{eq:grav-det}) corresponds to $h=1.$

It is now straightforward to put together the complete graviton determinant in (\ref{eq:grav-det}). The contribution of the spin-2 tensor determinant is given by setting $h=0$ in (\ref{eq:BHspin2det}), giving
\begin{eqnarray}
Z^{(1)}_{s=2, m_2=\pm1} &=& \prod_{\ell,\ell'=0}^\infty{1\over (1-q^{\ell + 2}\bar q^{\ell'}) (1-q^{\ell}\bar q^{\ell' + 2})}~.
\end{eqnarray}
Inserting this value and taking the ratio in (\ref{eq:grav-det}), we find
\begin{eqnarray}\label{eq:gravdetBH}
Z^{(1)}_{\rm grav} &=& \prod\limits_{\ell=0}^\infty\frac{1}{(1-q^{\ell+2})(1-\bar q^{\ell+2})}~.
\end{eqnarray}
This expression agrees with the results \cite{Maloney:2007ud,Giombi:2008vd}. 

There is a simple way to derive this final result without going through the process of constructing each determinant in (\ref{eq:grav-det}) explicitly. In particular, consider the quasinormal mode spectra in Tables \ref{table:BHQNMspin2} and \ref{table:BHQNMspin1}. Evaluating the conditions in these tables at $\Delta_{2} = 2$ and $\Delta_{1} = 3,$ we see that almost every spin-2 mode has a corresponding spin-1 ghost mode which satisfies the same condition. These modes will cancel when taking the ratio in the graviton determinant (\ref{eq:grav-det}). The only contributions which do not cancel are the spin-2 states at $p=0$ which satisfy
\begin{align}\label{eq:BHphys}
2i k_L &= 0 \qquad \text{for} \qquad m_2 >0~, \nn\\
2i k_R &= 0 \qquad \text{for} \qquad m_2 <0~.
\end{align}
As described in Appendix \ref{app:spin2Eucl}, one has to be careful with the Euclidean rotation of these states. In particular, as described in Appendix \ref{app:spin2Eucl}, in order to ensure that these Euclidean solutions are regular at the origin, the thermal quantum number $n$ should run only over a restricted set of values. Taking these restrictions into account and performing the sum we can directly recover (\ref{eq:gravdetBH}). This analysis demonstrates that the physical states that contribute to the graviton determinant come from either purely left-moving or purely right-moving states. This also explains the factorization in (\ref{eq:gravdetBH}), as the condition $2i k_L = 0$ yields the $\bar q$-dependent product in (\ref{eq:gravdetBH}) while the condition $2ik_R=0$ yields the remaining $q$-dependent part.

\section{ Quasinormal mode method:  Chiral boundary conditions }\label{sec:mixed}

We now move on to a further generalization of the DHS prescription, which will be the main focus of this article. The boundary conditions satisfied by quasinormal modes in the asymptotically AdS region correspond to Dirichlet boundary conditions. These are natural as they require fields to fall off in a prescribed way near the boundary such that small on-shell perturbations have a finite energy \cite{Breitenlohner1982}. However, certain types of fields in asymptotically AdS space-times allow for more general boundary conditions. For example, scalar fields with mass close enough to the Breitenlohner-Freedman bound can be quantized with Dirichlet or Neumann boundary conditions and still yield finite energy excitations \cite{Breitenlohner1982,Klebanov:1999tb}.%
\footnote{In \cite{DenefHartnollSachdev2010}, DHS do discuss Neumann conditions for these low-mass scalars, but only in the low-temperature limit.  Additionally their discussion is possible because when considering scalars in a non-rotating background, quasinormal modes simply map to Neumann-condition modes under $\tilde{\Delta}=d-\Delta$; as we discuss the mapping will be more complicated when fields with spin or backgrounds with rotation are considered.}
 Similarly, massless gauge fields, gravitons and higher spin fields can be quantized with Dirichlet or Neumann boundary conditions \cite{Giombi2013,Witten2003,Leigh2003,Marolf2006,Compere2008}. Below we will consider particular boundary conditions on the bulk metric which are a mixture of Dirichlet and Neumann boundary conditions. 

The goal of this section is to use a simple modification of the DHS argument to construct the one-loop determinant for the three dimensional graviton for   cases where certain components of the metric satisfy Neumann boundary conditions while others satisfy Dirichlet. As discussed in the previous section, the assumption that the one-loop determinant is meromorphic as a function of $\Delta$ implies that poles of the one-loop determinant occur whenever a quasinormal mode satisfies equation (\ref{omegarelation}). Our application of this method instead requires that we enforce Neumann boundary conditions for certain metric components. Our crucial working assumption is that these new boundary conditions will similarly quantize the frequency of ingoing (and outgoing) solutions such that poles of the determinant will now occur whenever the regularity condition
\be\label{omegarelationNeumann}
\tilde \omega_\star(\Delta_{\star,n})=\omega_n=2\pi i n T
\ee
holds.%
\footnote{For simplicity, in (\ref{omegarelationNeumann}) we reference the regularity condition for static backgrounds; for the non-static case one should use the more general condition discussed in section 2, which for rotating BTZ is given in (\ref{eq:omegarot}).} %
Here $\tilde \omega_\star(\Delta_{\star,n})$ refers to the quantized frequencies associated to ingoing (and outgoing) solutions which satisfy the prescribed Dirichlet-Neumann boundary conditions for each component at infinity. These will in general be different from the standard quasinormal frequencies. That the second equality in (\ref{omegarelationNeumann}) is unmodified relative to (\ref{omegarelation}) follows simply because the near-horizon analysis is independent of the asymptotic boundary conditions. In what follows we will refer to the frequencies $\tilde \omega_\star(\Delta_{\star,n})$ simply as quasinormal and also drop the tilde. In addition, we will utilize the more general prescription discussed in Section \ref{sec:Dscalar} appropriate to stationary but not necessarily static spacetimes.

We begin by reviewing the details of the various  boundary conditions for the metric that we will consider; then we move to a direct calculation of the one-loop determinant of the graviton (including its ghost contributions) following the philosophy discussed above.

\subsection{Chiral boundary conditions in AdS$_3$}\label{sec:bc}

We consider boundary conditions on metric fluctuations in asymptotically AdS$_3$ spacetimes which correspond to imposing Dirichlet or Neumann conditions on different components. In three dimensions it is natural to formulate a type of chiral boundary condition in which the left-moving components of the boundary metric are allowed to fluctuate (Neumann), whereas the right-moving components are held fixed (Dirichlet). 
Such chiral boundary conditions were initially proposed by Compere, Song and Strominger (CSS) in \cite{CompereSongStrominger2013a}; see also \cite{Troessaert2013b}. By additionally restricting the boundary metric to have purely left-moving coordinate dependence, CSS demonstrated that these boundary conditions modify the asymptotic symmetry algebra from a product of left and right-moving Virasoro algebras to a purely left-moving Virasoro plus $U(1)$ Kac-Moody algebra. 
Following \cite{CompereSongStrominger2013a}, the authors in \cite{Avery2014} realized that the left-moving coordinate dependence of the boundary metric in CSS could be relaxed. The resulting boundary conditions enhance the asymptotic symmetry algebra of CSS to an $sl(2,\RR)$ Kac-Moody; as such we will refer to these simply as ``$sl(2,\RR)$ KM'' boundary conditions.

For both boundary conditions, the starting point is pure AdS$_3$ gravity; the action is given by 
\be\label{act3d}
I_{3D}={1\over 16\pi G_3}\int d^3x \sqrt{-g^{(3)}}\left(R^{(3)}+{2}\right)~,
\ee
where the AdS radius is set to one.  We consider a class of backgrounds which have the following asymptotic behavior:
\bea\label{eq:ads3}
ds^2_{3D}&=& {dr^2\over r^2} - r^2(dt^+dt^- + h(t^+,t^-)(dt^+)^2)\cr
&&+ {4}G_3\mathfrak{m}\left(dt^- + f(t^+,t^-)dt^+ \right)^2 +{4 G_3} L(t^+,t^-) (dt^+)^2  +O(r^{-2})~.
\eea
Here $t^\pm=t\pm \phi$ with $\phi\sim \phi +2\pi$ and $\mathfrak{m}$ is a fixed constant. The Einstein equations impose some restrictions on the functions $h(t^+,t^-),$  $f(t^+,t^-),$ and $L(t^+,t^-)$; the remaining freedom on these functions is controlled by boundary conditions, which we will elaborate on below. 
In this notation, the BTZ black hole with mass $M$ and angular momentum $J$ corresponds to 
\be\label{eq:restrictions}
L(t^+,t^-)=L_0~,\quad h(t^+,t^-)=f(t^+,t^-)=0~,\quad  M=\mathfrak{m}+L_0~,\quad  J=\mathfrak{m}-L_0~,
\ee
where $L_0$ is  constant and $\mathfrak{m}>0$. Global AdS also falls into the restrictions in \eqref{eq:restrictions} upon setting $\mathfrak{m}=L_0=-1/G_3$.

\subsubsection{CSS boundary conditions}

The chiral boundary conditions of CSS \cite{CompereSongStrominger2013a} require that the boundary metric component $g_{++}$ depend only on the left-moving coordinate $t^+,$ such that
\begin{equation}\label{eq:hCSS}
h(t^+,t^-) = h(t^+)~.
\end{equation}
On-shell this condition  implies similar restrictions on the other metric functions: $f(t^+,t^-) = f(t^+)$ and $L(t^+,t^-)=L(t^+).$ Furthermore, the equations of motion also imply
\begin{equation}
f(t^+) = h(t^+) \equiv - \partial_+ P(t^+)~.
\end{equation}
The resulting metric has the asymptotic form
\bea\label{eq:ads3css}
{ds^2_{3D}}&=& {dr^2\over r^2} - r^2(dt^+dt^--\partial_+P(t^+)(dt^+)^2)\cr
&&+ {4G_3}\mathfrak{m}\left(dt^--\partial_+P(t^+)dt^+ \right)^2 +{4 G_3} L(t^+) (dt^+)^2  +O(r^{-2})~.
\eea

The $r$-dependence of the allowed fluctuations of the metric under diffeomorphisms becomes  
\begin{eqnarray}\label{eq:allowed3d}
&\delta g_{++}= O(r^2) ~, \quad \delta g_{+-} = O(1)~,\quad \delta g_{--}=O(r^{-2})~, &\nn\\
&\delta g_{r\pm} = O(r^{-3}) ~, \quad \delta g_{rr} = O(r^{-4}) ~.&
\end{eqnarray}
In other words, the allowed diffeomorphisms leave $\mathfrak{m}$ and the leading term of $g_{+-}$  fixed, whereas the functions $P(t^+)$ and $L(t^+)$ are allowed to fluctuate. 
Note that if we do not allow fluctuations of $\partial_+P(t^+)$, this analysis boils down to the holomorphic sector of the Brown-Henneaux boundary conditions. 

\subsubsection{$sl(2,\RR)$ KM boundary conditions}

A consistent extension of the CSS boundary conditions is to loosen the constraint $h(t^+,t^-) = h(t^+)$, while still holding $\mathfrak{m}$ fixed \cite{Avery2014}. In particular, by relaxing the falloff of $g_{r+}$ in \eqref{eq:allowed3d} such that 
\begin{equation}
\delta g_{r+} = O(r^{-1})~,
\end{equation}
instead of $O(r^{-3}),$ one finds that the Einstein equation constrains the $t^-$ dependence of the function $h(t^+,t^-)$ such that 
\begin{equation}
\partial_-\left(\partial_-^2 -16 G_3\mathfrak{m}\right)h(t^+,t^-) =0~,
\end{equation}
which is solved by
\begin{equation}\label{eq:hSL2}
h(t^+,t^-) = h(t^+) + g(t^+)e^{i N t^-} + \bar{g}(t^+)e^{-i N t^-}~,
\end{equation}
where $h(t^+),$ $g(t^+),$ and $\bar{g}(t^+)$ are arbitrary functions of $t^+$ and 
\begin{equation}
N^2 \equiv -16 G_3 \mathfrak{m}~.
\end{equation}
The remaining functions in the metric are constrained by the form of $h(t^+,t^-).$ In particular, $f(t^+,t^-)$ is now determined in terms of $h(t^+),$ $g(t^+),$ and $\bar{g}(t^+).$ $L(t^+,t^-)$ is similarly specified up to a function independent of $t^-,$ such that
\begin{equation}
L(t^+,t^-) = L(t^+) + \bar L(t^+,t^-)~,
\end{equation}
where $L(t^+)$ is an arbitrary periodic function of $t^+$ and $\bar L(t^+,t^-)$ is determined by $h(t^+),$ $g(t^+),$ and $\bar{g}(t^+).$ We refer the reader to \cite{Avery2014} for the full details. The important piece of information for us is that the radial falloff of the allowed diffeomorphisms for these boundary conditions are 
\begin{eqnarray}\label{eq:allowed3dSL2}
&\delta g_{++}= O(r^2) ~, \quad \delta g_{+-} = O(1)~,\quad \delta g_{--}=O(r^{-2})~, &\nn\\
&\delta g_{r+} = O(r^{-1}) ~, \quad \delta g_{r-} = O(r^{-3})~, \quad \delta g_{rr} = O(r^{-4}) ~.&
\end{eqnarray}

In the rest of this section we will use the DHS method to compute the one-loop determinant for both the CSS and $sl(2,\RR)$  KM boundary conditions. We will in particular focus on imposing the radial falloff conditions in \eqref{eq:allowed3d} and \eqref{eq:allowed3dSL2} and will then analyze the consistency with the chirality conditions on $h(t^+,t^-)$ given in \eqref{eq:hCSS} and \eqref{eq:hSL2}.

\subsection{Modified spin-2 determinant}\label{sec:ms2d}

In order to implement the DHS procedure for the boundary conditions discussed in section \ref{sec:bc}, we will first understand how the Neumann boundary conditions for $\delta g_{++}$ translate to boundary conditions on a massive spin-2 field and compute the corresponding determinant. For the massless case, we will also add a detailed discussion of the ghosts for both CSS and $sl(2,\RR)$ KM boundary conditions, highlighting subtleties that appear relative to the standard  scenario in section \ref{sec:gdd}.

\subsubsection{Massive spin-2 with chiral boundary conditions}\label{massivespin2chiral}

Our starting point is to specify how the chiral boundary conditions in (\ref{eq:allowed3d}) and (\ref{eq:allowed3dSL2}) translate to the boundary behavior of a massive spin-2 field $h_{\mu\nu}$ (as detailed in equations (\ref{eq:Rmplusin}) and (\ref{eq:Rmminusin}) of Appendix \ref{app:spin2}). In the following we will focus mainly on the tensor components along the boundary directions, and later on check that the remaining boundary conditions on the radial components are satisfied. Note that in the following,  the relevant extension to massive states of the mixed graviton boundary condition depends on the sign of the polarization, i.e. whether $m_2$ is positive or not. 

Near the boundary, a massive spin-2 field has the expansion 
\begin{eqnarray}\label{eq:spin2falloff}
h_{++} &\simeq& A_{++} r^{m_2+1}\left(1+\cdots\right) + C_{++} r^{-m_2-3}\left(1+\cdots\right)~, \\
h_{--} &\simeq& C_{--} r^{m_2-3}\left(1+\cdots\right) + A_{--} r^{-m_2+1}\left(1+\cdots\right)~,
\end{eqnarray}
where for conciseness we are only considering the relevant components to understand the chiral boundary conditions. It is worth mentioning that, according to the standard AdS/CFT dictionary, for $m_2=1$, $A_{++}$ acts as the source for the right-moving stress tensor $T_{--}$, whereas for $m_2=-1$, $A_{--}$ is the source for $T_{++}.$ However, the coefficients $C_{ij}$ do not act as the corresponding vacuum expectation values. Instead, for $m_2=1$, $A_{--}$ is the vev for the right-moving stress tensor $\langle T_{--}\rangle$ and vice versa for $m_2=-1.$ 

Given a boundary condition on a single component, the others are fixed by the first-order equations (\ref{eq:firstorderspin2}), so we only need to specify the behavior of a single component. For Brown-Henneaux, which are fully Dirichlet boundary conditions, we simply require that the source terms vanish, {\it i.e.}
\begin{equation}
{\rm Dirichlet ~B.C.:}\quad A_{++} = 0 \,\,\, \text{for}\,\,\,  m_2>0~, \qquad  \text{and} \qquad  A_{--} = 0 \,\,\,  \text{for}\,\,\, m_2<0~.
\end{equation}
To implement chiral boundary conditions we require that metric perturbations, $\delta g_{\mu\nu} = h_{\mu\nu}$ with $|m_2|=1,$ have right-moving components that fall off faster than a constant with
\begin{equation}
\delta g_{--} \sim o(r^0)~,
\end{equation}
while allowing for $\delta g_{++}$ to grow near the boundary. Comparing to the behavior in (\ref{eq:spin2falloff}), the natural extension of these boundary conditions away from the massless value corresponds to
\begin{equation}
{\rm Chiral~ B.C.:} \quad A_{--} = 0 \,\,\, \text{for}\,\,\,  m_2>0~, \qquad  \text{and} \qquad  A_{--} = 0 \,\,\,  \text{for}\,\,\, m_2<0~.
\end{equation}
For $m_2<0,$ this is the same boundary condition as in the standard Dirichlet situation. However, for $m_2 > 0,$ we are imposing Neumann boundary conditions, as we are holding $\langle T_{--}\rangle$ fixed and allowing the source to fluctuate.

The quasinormal modes\footnote{Perhaps these should not be referred to as ``normal" anymore as the mode functions are not square-normalizable at the boundary for $\Delta_2>2$. However, while acknowledging this abuse of terminology, we will still refer to these as quasinormal modes.} associated with these boundary conditions are derived in Appendix \ref{app:mixedQNM}. The end result for the quasinormal spectrum with chiral boundary conditions is given in the following table:

\noindent\begin{minipage}{\linewidth}
	\begin{center}
		\begin{tabular}{|c|c|c|}
			\hline
			& ingoing & outgoing\\
			\hline
			$m_2 > 0$	& $\begin{matrix}\, 2ik_R = 2p - \Delta_2  \,\,\, \\
			\,2ik_L =  2p - \Delta_2 + 4  \,\,\,\end{matrix}$&  $\begin{matrix}\, 2ik_R = -(2p - \Delta_2)  \,\,\, \\
			\,2ik_L = -(2p - \Delta_{2} + 4)  \,\,\,\end{matrix}$\\ 
			\hline
			$m_2 < 0$ & $\begin{matrix}\, 2ik_R = 2p + \Delta_{2} - 2 \,\,\, \\
			\,2ik_L =  2p + \Delta_{2} + 2  \,\,\, \end{matrix}$ & $\begin{matrix}\, 2ik_R = -(2p + \Delta_{2} - 2)  \,\,\, \\
			\,2ik_L =  -(2p + \Delta_{2} + 2)  \,\,\,\end{matrix}$\\
			\hline
		\end{tabular}\captionsetup{justification=raggedright}\captionof{table}{Spin-$2$ quasinormal mode spectrum after imposing chiral boundary conditions. When $\Delta_{2}=2$ these correspond to the symmetric, transverse, traceless graviton spectrum. Each condition on $k_R$ or $k_L$ labels a distinct eigenmode and the range of $p$ is over all non-negative integers.}\label{table:mixedQNMspin2}
	\end{center}
	\vspace{.2em}
\end{minipage}
Here we have again organized the modes into ``ingoing" and ``outgoing" based on their behavior at the horizon. Notice that since  the $m_2<0$ states still satisfy Dirichlet boundary conditions, the quasinormal modes in this sector are precisely the same as they were in the previous section. It is also interesting to note that the conditions on the new $m_2>0$ states in table \ref{table:mixedQNMspin2} are the same conditions as those on the $m_2 <0$ states upon sending $\Delta_2 \rightarrow 2-\Delta_2.$ This suggests that both sets of states have the same chirality and we will see this feature in the final result for the one-loop determinant. Finally, the swapping of $\Delta_2$ with $2-\Delta_2$ for $m_2>0$ naturally follows from the alternative (Neumann) quantization of these states.

Enumerating the Euclidean solutions in this case is very similar to the situation with Dirichlet boundary conditions. We summarize the conditions on the Euclidean spectrum in the following table.

\noindent\begin{minipage}{\linewidth}
	\begin{center}
		\begin{tabular}{|c|c|}
			\hline
			$m_2>0$ & $m_2<0$ \\
			\hline
			$\begin{matrix}\, 2p + 2 - \Delta_2 + |n-2| + ik_\Phi(n,k) = 0  \,\,\, \\
			\,  2p + 2- \Delta_2 + |n+2| - ik_\Phi(n,k) = 0   \,\,\,\end{matrix}$&  
			$\begin{matrix}\, 2p + \Delta_2 + |n-2| + ik_\Phi(n,k) = 0   \,\,\, \\
			\,2p + \Delta_2 + |n+2| - ik_\Phi(n,k) = 0 \,\,\,\end{matrix}$\\ 
			\hline
		\end{tabular}\captionsetup{justification=raggedright}\captionof{table}{Conditions satisfied by Euclidean solutions with chiral boundary conditions. Each solution satisfies one of the conditions listed. Here $k_\Phi(n,k)$ is given in equation (\ref{eq:kPhi}). In this table, $p$ runs over all non-negative integers, whereas $n$ and $k$ run over all integers.}\label{table:mixedEuclCond}
	\end{center}
	\vspace{2em}
\end{minipage}
We can now compute the contribution to the one-loop determinant from all of the $m_2>0$ states in Table \ref{table:mixedEuclCond}. This gives
\begin{equation}
Z^{(2)}_{m_2>0,\text{Neumann}} = \prod_{\ell,\ell'=0}^\infty{1\over (1-q^{\ell + h'+2}\bar q^{\ell' + h'})}~,
\end{equation}
where $h' =  -{\Delta_2}/{2}.$ Putting this together with the Dirichlet result for $m_2 <0,$ we have
\begin{equation}\label{eq:spin2detmixed}
Z^{(1)}_{s=2,\text{chiral}}(\Delta_2) = \prod_{\ell,\ell'=0}^\infty{1\over (1-q^{\ell + h'+2}\bar q^{\ell' + h'})(1-q^{\ell+h+2}\bar q^{\ell' + h})}~.
\end{equation}

Before moving on, we would like to comment on the $e^{\rm Pol(\Delta_2)}$ factor that we have dropped in the expression for the one-loop determinant above. In this case, the determination of this factor is potentially subtle. In particular, consider the $\Delta_2\rightarrow \infty$ limit of (\ref{eq:spin2detmixed}). For the second factor, which arises from the $m_2<0$ Dirichlet contribution, taking $\Delta_2\rightarrow \infty$ is straightforward. However, in the first (Neumann) factor it appears that one should instead take $\Delta_2 \rightarrow -\infty$ in order for the limit to commute with the product over $(\ell,\ell').$ Perhaps this could be expected to be the case since the alternative quantization is naturally phrased in terms of $\Delta_- = 2 - \Delta_2,$ and taking $\Delta_- \rightarrow \infty$ corresponds to $\Delta_2\rightarrow - \infty.$ A proper understanding of heat kernel techniques for the chiral boundary conditions considered here would likely address this issue. Since this does not affect the pole structure of the one-loop determinant, we leave such an analysis for future work.

\subsection{The graviton one-loop determinant}\label{sec:mdngraviton}

We now construct the graviton one-loop determinant for CSS and $sl(2,\RR)$ KM boundary conditions from the results for the massive spin-2 determinants. As in \eqref{eq:grav-det}, we need to evaluate
\be\label{eq:grav11}
Z^{(1)}_{\rm grav} = \left(\frac{\det_{\rm T}(-\nabla^2 + 2/L^2)}{\det_{\rm STT}(-\nabla^2 - 2/L^2)}\right)^{1/2}~.
\ee 
The denominator is straightforward to obtain from the massive case: we just set $\Delta_2=2$ in  \eqref{eq:spin2detmixed}. The numerator, which is the contribution from the ghost fields, is more delicate: results vary depending on whether we impose the boundary conditions on the vector field itself or on the metric perturbation they induce as we will show in the following.

\subsubsection{The ghost contribution} 

It turns out that we have already determined most of the ghost contribution to (\ref{eq:grav-det}). In particular, as detailed in Appendix \ref{app:spin1}, for the ghost fields the standard Dirichlet boundary conditions are already consistent with the new chiral boundary conditions. This means that the states in Table \ref{table:BHQNMspin1} will contribute just as they had in the case with Dirichlet boundary conditions. There are, however, several additional sets of quasinormal modes which satisfy chiral boundary conditions, but not Dirichlet. These are given in Table \ref{table:Neumannspin1modes}. As we will discuss, whether or not we include these extra modes will play an important role in what follows.

\begin{table}
	\begin{center}
		\begin{tabular}{|c|c|}
			\hline
			ingoing& outgoing \\
			\hline
			$ - 2ik_R + \Delta_1 -1 = 0$ &  $2ik_R + \Delta_1 - 1 = 0$ \\
			$ - 2ik_R - (\Delta_1 -1) = 0$ &  $2ik_R - (\Delta_1 - 1) = 0$\\ 
			\hline
			$-2 i k_R=0$ & $2i k_R=0$ \\
			\hline 
		\end{tabular}\captionsetup{justification=raggedright}\caption{\small Additional spin-1 ghost states that are consistent with the Neumann conditions on $g_{++},$ but are not contained in the Brown-Henneaux states. The first two lines correspond to new $m_1>0$ states, whereas the $k_R=0$ states arise both in the $m_1>0$ and $m_1<0$ sectors. }\label{table:Neumannspin1modes}
	\end{center}
\end{table}

As explained in Appendix \ref{app:spin1}, when considering the spin-1 states at the value of the ghost mass, corresponding to $\Delta_1=3,$ there are special states that appear in the second and third rows of Table \ref{table:Neumannspin1modes} that are actual zero modes of the ghost Laplacian, which locally satisfy the Killing equation.%
\footnote{Here ``actual zero modes'' refers to modes with zero eigenvalue in the determinant when $\Delta$ is tuned to its physical value. Their contribution to the path integral yields a prefactor which scales with the number of such zero modes, which we are neglecting.} %
In particular, these occur for $|k_E|= 1$ in the Euclidean solutions for the $m_1=2$ states in the second row as well as the $m_1=-2$ states in the third row of Table \ref{table:Neumannspin1modes}.  Since these are zero modes of the ghost Laplacian they will only appear in ${\rm Pol}(\Delta_1)$ but not in the poles of $Z^{(1)}$.
Taking into account all of the ghost states that induce a pole, we find
\begin{eqnarray}\label{eq:ghostdetmixed}
Z^{(1)}_{s=1,\text{chiral}} &=&\prod_{\ell,\ell'=0}^\infty{1\over (1-q^{\ell + 2}\bar q^{\ell' + 1})(1-q^{\ell + 1}\bar q^{\ell' + 2})} \nn \\ 
&& 
\times\prod_{\ell=0}^\infty{1\over (1-q^{\ell + 1}\bar q) (1-q^{\ell+1})(1-q^{\ell+1}\bar q^{-1})(1-q^{\ell+2})}~,
\end{eqnarray}
where the first product corresponds to the contribution which is also included in the Brown-Henneaux analysis and the second product is from the new states in Table \ref{table:Neumannspin1modes}.

Before proceeding, there is an important point to make regarding which ghost modes we allow to contribute to the physical graviton determinant. The ghost is a vector field $V_\mu$ that induces the gauge transformation
\be\label{eq:gghostv}
\delta g_{\mu\nu} = \nabla_{\mu} V_\nu + \nabla_\nu V_\mu~.
\ee
 When one imposes Dirichlet boundary conditions on $V_\mu$, one finds that {\rm all} the induced metric fluctuations by such $V_\mu$   falloff faster than the Brown-Henneaux boundary conditions \eqref{eq:bhbc}. However, allowing for Neumann boundary conditions for $V_\mu$ introduces the possibility that the ghost eigenfunctions will generate metric variations which are of the same order as the allowed falloffs in (\ref{eq:allowed3d}) and (\ref{eq:allowed3dSL2}): these are the states the second line of \eqref{eq:ghostdetmixed}, which correspond to the modes in Table \ref{table:Neumannspin1modes}. Whether or not we choose to keep these modes depends on how we implement boundary conditions:
 \begin{enumerate}
\item We could impose that $V_\mu$ cannot induce a metric fluctuation \eqref{eq:gghostv} as leading as those allowed by the asymptotic symmetry group;
\item Or we could impose Neumann boundary conditions on $V_\mu$, and hence allow for large induced metric fluctuations compatible with  the asymptotic symmetry group. 
  \end{enumerate}
In what follows we will be agnostic about these ghost contributions and present the determinant for both situations. We will elaborate on the meaning of the subsequent results when we discuss the holographic interpretation of the various boundary conditions in Section \ref{sec:holography}. 

\subsubsection{$sl(2,\RR)$ KM determinant}

In order to distinguish between the $sl(2,\RR)$ KM boundary conditions and the CSS boundary conditions, we need to consider the boundary falloff of the $\delta g_{r+}$ components. For the $sl(2,\RR)$ KM boundary conditions in \eqref{eq:allowed3dSL2}, the boundary condition on $\delta g_{r+}$ coincides with the generic behavior of a solution with Neumann conditions on $\delta g_{++}$ and so all of the spin-$2$ states enumerated above contribute to the $sl(2,\RR)$ KM determinant. Next, for the ghost fields, if we require that \eqref{eq:gghostv} is subleading relative to \eqref{eq:allowed3dSL2}, only the first line of \eqref{eq:ghostdetmixed} contributes. Combining these two contributions in \eqref{eq:grav11}, we find the following result for the graviton determinant
\begin{eqnarray}\label{eq:gravdetSL2a}
Z^{(1)'}_{sl(2,\RR)} &=& \prod_{\ell,\ell'=0}^\infty
{(1-q^{\ell + 2}\bar q^{\ell' + 1})\over (1-q^{\ell+2}\bar q^{\ell'})}{(1-q^{\ell + 1}\bar q^{\ell' + 2})\over(1-q^{\ell + 1}\bar q^{\ell' -1})}
\nn\\
&=&\prod_{\ell=0}^\infty{{1}\over (1-q^{\ell + 1}\bar q) (1-q^{\ell+1})(1-q^{\ell+1}\bar q^{-1})(1-q^{\ell+2})}~.
\end{eqnarray}
Here we have written the determinant with a prime to emphasize that we have not included any of the ghost contributions which induce metric fluctuations of the same order as those allowed by \eqref{eq:allowed3dSL2}. 

Now, let us consider what happens when we include the ghost degrees of freedom that grow near the boundary. To do this we must simply keep all of the terms in  \eqref{eq:ghostdetmixed}. This will precisely cancel the expression in (\ref{eq:gravdetSL2a}) and we arrive at the final result
\begin{equation}\label{eq:gravdetSL2}
Z_{sl(2,\RR)}^{(1)} = 1.
\end{equation}
We will comment on the interpretation of this result and the expression in (\ref{eq:gravdetSL2a}) in Section \ref{sec:holography}.

\subsubsection{CSS determinant}

We will now move on to construct the determinant for CSS boundary conditions \eqref{eq:allowed3d}. Relative to the $sl(2,\RR)$ KM case, we have the more stringent restriction 
\begin{equation}\label{eq:CSSg+r}
\delta g_{+r} \sim O(r^{-3})~.
\end{equation}
In appendix \ref{app:gr+}, we find that the condition in \eqref{eq:CSSg+r}, along with the other CSS conditions in (\ref{eq:allowed3d}), are generically only satisfied if we impose simultaneously that the leading term in $h_{LL}$ and $h_{RR}$ vanish. In addition, there is a special state with $k_R=0$ for which only the leading term in $h_{RR}$ must vanish in order satisfy all of the CSS conditions. In other words, for  $m_2>0$ spin-$2$ states,\footnote{Recall that the $m_2<0$ states are insensitive to the current discussion as they are required to satisfy Dirichlet boundary conditions.} the spectrum is given by the intersection of states in Table \ref{table:mixedQNMspin2} with those in Table \ref{table:BHQNMspin2} together with the $k_R=0$ state. When $\Delta_2 = 2$, we find that the resulting states are
\be\label{eq:kr1}
2ik_R=2p+4~,\qquad {\rm with} \quad p=-2,0,1,2,3,\ldots~,
\ee
and
\be\label{eq:kl1}
ik_L=p~,\qquad {\rm with} \quad p=1,2,3,\ldots~.
\ee
The restricted set of states in \eqref{eq:kr1}, compared to those in Table \ref{table:mixedQNMspin2}, means we should remove from \eqref{eq:gravdetSL2a} a factor of
\begin{equation}
\prod\limits_{\ell=0}^{\infty}{1\over (1-q^{\ell+1}\bar q)(1-q^{\ell+1}\bar q^{-1})}~.
\end{equation}
 That these states are removed could have been anticipated as they correspond to fluctuations of the boundary metric function $h(t^+,t^-)$ that  violate the chirality condition in \eqref{eq:hCSS}.  The condition \eqref{eq:kl1} is as stated in Table \ref{table:mixedQNMspin2} for $\Delta_2=2$, so no further modification of the spin-$2$ contribution to \eqref{eq:gravdetSL2a} is required. It is interesting to note that the $k_L= 0$ states are not contained in the CSS spectrum: these states give the $1/(1-\bar q^{\ell+2})$ in the standard Brown-Henneaux result (\ref{eq:gravdetBH}) and it is nice to see that the Neumann conditions naturally exclude these.

Finally, we need to consider the ghost contribution to the determinant. In this case, the result is simple. None of the new states in the first two lines of Table \ref{table:Neumannspin1modes} generate metric variations that satisfy \eqref{eq:CSSg+r}. Furthermore, the $k_R=0$ states generate metric variations which falloff precisely as fast as the allowed metric boundary conditions. Since we would like to define the modes which saturate the boundary falloffs in (\ref{eq:allowed3d}) as the physical boundary gravitons, we should in addition exclude the $k_R=0$ states from the determinant. This means that none of the terms in the second product in \eqref{eq:ghostdetmixed} contribute: the ghost determinant for CSS is just given by  the first line. The net sum of these restrictions yields 
\begin{equation}\label{eq:gravdetCSS}
Z_{\rm CSS}^{(1)'} = \prod\limits_{l=0}^\infty \frac{1}{(1-q^{\ell+1})(1-q^{\ell+2})}
\end{equation}
as the final result for the CSS determinant. 

As in the Dirichlet case, there is a simple way of deriving the result in \eqref{eq:gravdetCSS} without first going through the full computation of spin-1 and spin-2 determinants separately. In a similar fashion to the discussion around \eqref{eq:BHphys} for the Brown-Henneaux states, when $\Delta_2=2$ and $\Delta_1=3$ almost all of the quasinormal mode conditions on the CSS spin-2 states are matched with conditions on ghost states with the exception of the two states
\begin{align}\label{eq:CSSphys}
2i k_R &= 0 \qquad \text{for} \qquad m_2 >0, \nn\\
2i k_R &= 0 \qquad \text{for} \qquad m_2 <0.
\end{align}
Comparing to the conditions in (\ref{eq:BHphys}), which yielded one sum over left-movers and another over right-movers, here we instead have two sums over left-movers only. This is why the final result only depends on $q.$

Notice also that there is a difference in the exponent of $q$ in the two factors in (\ref{eq:gravdetCSS}). The origin of this can be seen by noticing that the condition in (\ref{eq:CSSphys}) for $m_2>0$ corresponds to the state at the $p=1$ level in Table \ref{table:mixedQNMspin2}, as opposed to the $p=0$ level as is the case for the other states in (\ref{eq:BHphys}) and (\ref{eq:CSSphys}). As detailed in Appendix \ref{app:spin2}, regular Euclidean solutions at the $p=0$ and $p=1$ levels allow for only a restricted set of thermal frequencies. For $p=0,$ the excluded thermal frequencies lead to the shift of $\ell \rightarrow \ell+2$ in the exponents of $q$ and $\bar q$ in the graviton partition functions. For $p=1,$ the exponent is only shifted to $\ell+1,$ giving the additional $(1-q)^{-1}$ relative to the other cases. 

Finally, as in the $sl(2,\RR)$ KM case, one can in principle include the ghost states which induce metric variations that have radial falloffs on par with the boundary gravitons. In this case this amounts to including the $k_R=0$ modes in Table \ref{table:Neumannspin1modes}. Doing so, we again find that the one-loop determinant trivializes
\begin{equation}\label{eq:gravdetCSSgauged}
Z^{(1)}_{\rm CSS} = 1.
\end{equation}
We will elaborate on the meaning of the one-loop determinants computed here in the next section.

\section{Holographic interpretation}\label{sec:holography}

We will now gather all the determinants we have evaluated in previous sections and discuss their holographic interpretation. Our aim is to highlight how to write the determinants as traces over unitary representations of the dual theory. This excludes the interpretation of the entire function $e^{{\rm Pol}(\Delta)}$; the emphasis is only on the interpretation of the pole structure of $Z^{(1)}(\Delta)$. 

\subsection{Standard boundary conditions}

This subsection will serve mostly as review, since the interpretation was already discussed in \cite{Maloney:2007ud,Giombi:2008vd}. The punchline in this case is that for standard (Dirichlet) boundary conditions we can interpret each determinant as the character of the two dimensional conformal group. This is in perfect agreement with the statement that these boundary conditions are precisely those behind AdS$_3$/CFT$_2$. The discussion here should be contrasted with the results in the following subsections.  
\begin{description}
	\item[Real Scalars:] In section \ref{sec:Dscalar} we found that the one-loop determinant of a real scalar field is
	\bea\label{eq:Zs}
	\log Z^{(1)}(\Delta)=\log \det (-\nabla^2+m^2)^{-1/2} 
	&=&\log \prod_{\ell,\ell'=0}{1\over 1 - q^{\Delta/2+\ell}\bar{q}^{\Delta/2+\ell'}}~.
	\eea
	As in \cite{Giombi:2008vd}, it is useful to digest a bit this answer and view it as a trace, i.e. we want to interpret \eqref{eq:Zs} as
	\be
	{\rm Tr} \,q^{L_0}\bar q^{\bar L_0}~.
	\ee 
	where $L_0$ and $\bar L_0$ are each elements of an $sl(2)$ algebra, which we parametrize as
	\be
	[L_i,L_j]=(i-j)L_{i+j} ~,\qquad i,j=-1,0,1;
	\ee
	and similarly for $\bar L_i$. Rewriting \eqref{eq:Zs} as
	\be\label{eq:Zs1}
	\prod_{\ell,\ell'=0}{1\over 1 - q^{\Delta/2+\ell}\bar{q}^{\Delta/2+\ell'}} = \prod_{\ell\ell'} \sum_{n=0}^\infty q^{n(\Delta/2+\ell)}\bar{q}^{n(\Delta/2+\ell')}
	\ee
	makes the holographic interpretation of \eqref{eq:Zs} quite straight forward. The scalar field of mass $m^2=\Delta(\Delta-2)$ corresponds to a primary in a CFT$_2$ with conformal dimensions $(\Delta/2,\Delta/2)$. We denote  a single particular  state associated to the scalar field as $|h,h\rangle$, with $\Delta=2h$; multi-particle states correspond to multiple insertions of the operator at the origin.  The state $|h,h\rangle$ is annihilated by $L_1$ and $\bar L_1$ and a descendent of conformal weight $(\ell+h,\ell'+h)$ is given by
	\be
	L_{-1}^\ell \bar L_{-1}^{\ell'}|h,h\rangle ~, \qquad \ell,\ell'\geq 0~.
	\ee	
	The interpretation of the partition in \eqref{eq:Zs1} is now clear: the contribution for fixed $(\ell,\ell')$ corresponds to the trace of multi-particle configurations of a given descendent state of $|h,h\rangle$. Note that the states of the scalar operator are organized as  a character of $sl(2)\times sl(2)$; the full Virasoro algebra will only be evident for the graviton determinant.

	\item[Massive Vectors \& Tensors:] The result for a massive vector field in AdS$_3$ was derived in \eqref{eq:BHspin1det}
	\bea\label{Adet}
	Z^{(1)}_{s=1} &=&  \prod_{\ell,\ell'=0}^\infty{1\over (1-q^{\ell +h+1}\bar q^{\ell' + h}) (1-q^{\ell +h}\bar q^{\ell' + h + 1})}~,
	\eea
	and for a massive spin-2 field we found in \eqref{eq:BHspin2det}
	\be\label{Tdet}
	Z^{(1)}_{s=2} = \prod_{\ell,\ell'=0}^\infty{1\over (1-q^{\ell +h+2}\bar q^{\ell' + h}) (1-q^{\ell +h}\bar q^{\ell' + h + 2})}~.
	\ee
	The conformal dimension is $\Delta_{s} = 2 h + s$, and  a massless field has $h=0$. 
	
	The trace interpretation of \eqref{Adet}  and \eqref{Tdet} works very similarly to the scalar case. The only difference is that the vector and tensor have two polarization states: $(h,h+s)$ and $(h+s,h)$. Additionally, for each polarization state we have a tower of descendants of $sl(2)\times sl(2)$ and the multi-particle state configurations.  
	
	\item[Graviton:] We now turn to the determinant of the graviton with standard (Dirichlet) boundary conditions; the answer in \eqref{eq:gravdetBH} reads
	\begin{eqnarray}\label{eq:GD}
	Z^{(1)}_{\rm grav} = \prod\limits_{\ell=2}^\infty\frac{1}{(1-q^{\ell})(1-\bar q^{\ell})}~.
	\end{eqnarray}	
	Here the interpretation deviates slightly from our previous examples. Interestingly, because it is dual to the CFT stress tensor, the graviton captures the full structure of the Virasoro group, in contrast to the global $sl(2)\times sl(2)$ as seen above. If we denote the vacuum state as $|0\rangle$, the one-loop determinant \eqref{eq:GD} is counting descendants
	\be
	L_{-n_1}\cdots L_{-n_i} \bar L_{-n'_1}\cdots \bar L_{-n'_j} |0\rangle ~, \quad n_i,n'_j>1~,
	\ee
	where
	\be
	[L_n,L_m]=(n-m)L_{m+n} +\frac{c}{12} (n^3-n)\delta_{m+n}~,
	\ee
	and similarly for $\bar L_n$.
	Note that the vacuum state is annihilated by $L_{-1}$ and  $\bar L_{-1}$ and hence the product in \eqref{eq:GD} is from $\ell=2$. This is completely compatible with the results of Brown-Henneaux \cite{Brown:1986nw}: with Dirichlet boundary conditions, the spectrum of gravitational  solutions is organized with respect to two copies of the Virasoro algebra with central charge $c=3\ell_{\rm AdS}/2G_3$.  
	
	The determinant was evaluated in the BTZ background, however we are interpreting the resulting product formula as a vacuum character, which we would attribute to thermal AdS. The reason is simple: the Euclidean solutions, BTZ and thermal AdS, are indistinguishable since both are a quotient of Euclidean AdS$_3$ \cite{BanadosHenneauxTeitelboimEtAl1993,CarlipTeitelboim1995a,MaldacenaStrominger1998a}. It is only the Lorentzian continuation that makes them physically distinct: the Wick rotation to Lorentzian signature identifies if either a timelike or spatial cycle is contractible versus non-contractible in the Euclidean torus. This Wick rotation in addition changes the role of $\tau$ in the geometry; if for BTZ we have complex structure $\tau$ then thermal AdS corresponds to $-1/\tau$. In the language of the dual CFT$_2$ this is expected from modular invariance: the states at high temperature (BTZ) are related to low temperature excitations (thermal AdS).     
	
\end{description}

\subsection{Chiral boundary conditions}

In the following we will give an interpretation of the graviton one-loop determinants which involved chiral boundary conditions. There are two types of falloff that we considered in section \ref{sec:mixed}. As we will see below their interpretation is dramatically different and will depend on how we choose to implement the ghost determinant with Neumann boundary conditions.

\subsubsection{CSS boundary conditions} 
\label{sec:CSShol}

The analysis of the asymptotic symmetry group with boundary conditions \eqref{eq:allowed3d} suggests that its dual description should be in terms of a warped conformal theory (WCFT). These theories all have the following symmetry features: Given a coordinate system $(x^+, x^-)$, a WCFT is classically invariant under the transformations
\be\label{a1}
x^{+}~ \to~ x^+ + g(x^-)  ~,\qquad  x^- ~\to ~f(x^-) ~,
\ee
where $f$ and $g$ are arbitrary functions. The algebra of charges associated to these transformations is
\bea
\label{eq:canonicalgebra}
[ L_n, L_{m}] &=&(n-m) L_{n+m}+\frac{c}{12}n(n^2-1)\delta_{n+m}~, \cr
[ L_n, P_{m}] &=&-m  P_{m+n}~, \cr
[ P_n, P_{m}] &=&k \frac{n}{2}\delta_{n+m}~,
\eea
which is a Virasoro-Kac-Moody algebra with central charge $c$ and level $k$. Here $P_n$ generate diffeomorphisms along $x^+$ in \eqref{a1} \cite{HofmanStrominger2011,DetournayHartmanHofman2012}: this is what distinguishes a WCFT from other realisations of the Virasoro-Kac-Moody algebra. It is important to stress that this is a \emph{chiral} algebra (there is no $\bar L_n$ sector), and this chirality will be crucial as we interpret our results.

To start, let us review a few facts about unitary representations of \eqref{eq:canonicalgebra}; the discussion here is based on results in \cite{DetournayHartmanHofman2012,CastroHofmanSarosi2015}. A primary state is defined as a state $|p,h\rangle$ that is an eigenstate of the zero modes
\be
P_0|p,h\rangle = p |p,h\rangle~, \qquad L_0|p,h\rangle=h|p,h\rangle~,
\ee
and is annihilated by $(L_{n},P_n)$ with $n>0$. Descendants are created by acting with $L_{-n}$ and $P_{-n}$ ($n>0$). The trace that counts the descendants of a single primary reads  
\be\label{eq:x1}
{\rm Tr} \le( q^{L_0}\bar q^{ P_0} \ri)= q^h \bar q^p   \phi(q)^{-1} \chi_{ h}(q) ~.
\ee
The descendants created by acting with $P_{-n}$'s on   $| h,p\rangle $ are accounted by the Euler phi function
\be
\phi(q)=\prod_{n=1}^{\infty}(1-q^n)~,
\ee
while the descendants arising from the action of $L_{-n}$'s are counted by an ordinary Virasoro character, $\chi_{h}(q)$,
with central charge $c$. We note that a descendant state does \emph{not} shift the eigenvalue of $P_0$ and hence the character in \eqref{eq:x1} is holomorphic in $q$ (up to the overall dependence of $\bar q^p $). Finally, the global part of \eqref{eq:canonicalgebra} is simply $sl(2)\times u(1)$: characters of this algebra will be just labelled by the $sl(2)$ piece.

With this background, we can now proceed to interpret the determinants we evaluated in section \ref{sec:mdngraviton}. For the graviton we found in \eqref{eq:gravdetCSS} the following
\begin{equation}\label{eq:gravdetCSS1}
Z^{(1)}_{\rm grav,CSS} = \prod_{\ell'=1}^{\infty}{1\over (1-q^{\ell'})}\prod_{l=2}^\infty \frac{1}{(1-q^{\ell})}~.
\end{equation}
This is in perfect agreement with \eqref{eq:x1} when the primary state is the vacuum state: the first product is counting the $P_{-n}$ descendants, and the second product is the Virasoro character for $c>1$ with the $L_{-1}$ state removed.%
\footnote{\label{qhbarqPfootnote}Here we are just focusing on the pole structure of the one-loop contribution; the classical piece of the action and $e^{\rm Pol(\Delta)}$ will capture the $q^h\bar q^P$ piece of the trace.}  %
It is remarkable that the final result is holomorphic as expected from \eqref{eq:x1}. We stress that in a WCFT, suitable warped modular transformations also relate thermal AdS and BTZ  \cite{DetournayHartmanHofman2012,SongWenXu2017}. This relationship explains why we obtain a vacuum character when evaluating the determinant on BTZ.

As for Dirichlet boundary conditions, it is also interesting to interpret the determinant of massive fields. For instance, the massless spin-2 determinant with CSS boundary conditions is given by
\begin{equation}
Z^{(1)}_{s=2,{\rm CSS}}(\Delta_2) =  \prod_{\ell,\ell'=0}^\infty{1\over (1-q^{\ell + h'+2}\bar q^{\ell' + h'})(1-q^{\ell+h+2}\bar q^{\ell' + h})}~.
\end{equation}
Note that this determinant does not fit with the global part in \eqref{eq:canonicalgebra}: the $\bar q$ dependence cannot be accounted for by the Virasoro-Kac Moody algebra. The graviton respects the symmetries expected from ASG analysis, but matter in this theory is not organized by the same principle. It is possible to obtain a result compatible with $sl(2)\times u(1)$ representations, but this requires fixing the quantum number associated to $P_0$ in the quasinormal mode spectrum. We find this requirement strange; for the graviton in \eqref{eq:gravdetCSS1} we did not have to implement such a constraint.

It is worthwhile to compare our result with prior literature. The original derivations \cite{CompereSongStrominger2013a,Troessaert2013b} do not obtain \eqref{eq:canonicalgebra}; they obtain a \emph{non-canonical} form of the algebra where the commutator of $L_n$ and $P_n$ is shifted and the level depends on the vev of $P_0$ (which is $\mathfrak{m}$ in \eqref{eq:ads3css}). However,  \cite{SongWenXu2017} argue that  there is a non-local transformation that  brings the algebra to the form \eqref{eq:canonicalgebra}, where $k$ is independent of state, and modular invariance in the WCFT is restored (since $P_0$ can now vary). Our derivations are compatible with \eqref{eq:canonicalgebra} and modular invariance, hence we are indirectly justifying the non-local transformation advocated in  \cite{SongWenXu2017}.

Finally, we should discuss the interpretation of the result in \eqref{eq:gravdetCSSgauged}, where we have included the ghost states that are growing near the boundary. The natural interpretation of this result is in terms of a two-dimensional theory of induced gravity, where the additional ghost states represent the gauge redundancies in the boundary theory. However, one should not think of this as gauging the symmetries of a unitary WCFT, but instead simply in terms of 2d quantum gravity in a chiral light-cone gauge \cite{CompereSongStrominger2013}. This is in seeming conflict with the WCFT interpretation of the CSS boundary conditions that we have just discussed since in order to gauge the Virasoro $U(1)$ KM symmetry the level $k$ must be negative. The ability to treat the asymptotic symmetries as either global or gauge symmetries appears to be related to the fact that in gravity one finds the non-canonical form of the WCFT algebra with the WCFT description only emerging once one allows for the non-local transformations described in \cite{SongWenXu2017}. It would be worthwhile to understand this point more completely. As we will discuss in the next section, the interpretation in terms of induced gravity will be much more transparent in the theory with $sl(2,\RR)$ KM boundary conditions.

\subsubsection{$sl(2,\RR)$ KM boundary conditions}\label{sec:sl2hol}

The $sl(2,\RR)$ KM boundary conditions are distinguished from those of CSS by relaxing the chirality condition on the boundary metric in (\ref{eq:hCSS}). The asymptotic symmetry analysis for these boundary conditions was performed in \cite{Avery2014}, where the asymptotic symmetry algebra was shown to be a semidirect sum of a Virasoro and an $sl(2,\RR)$ KM current algebra. The generators satisfy the following commutation relations
\begin{eqnarray}
\label{eq:sl2KMalgebra}
[ L_n, L_{m}] &=&(n-m) L_{n+m}+\frac{c}{12}n(n^2-1)\delta_{n+m}~, \cr
[ L_n, J^a_{m}] &=&-m  J^{a}_{m+n}~, \cr
[ J^a_n, J^b_{m}] &=&f^{ab}{}_c J^{c}_{m+n} - k \frac{m}{2}\eta^{ab}\delta_{n+m}~,
\end{eqnarray}
where $f^{ab}{}_c$ are the structure constants of $sl(2,\RR)$ and  $\eta^{00} = -1,$ $\eta^{+-} = 2,$ while the other components of the metric $\eta^{ab}$ vanish. Finally, the level $k$ of the current algebra is determined by the central charge and is given by 
\begin{equation}
k = \frac{c}{6} = \frac{1}{4G_3}.
\end{equation}
Generically, $k$ and $c$ do not have to be related: it is a feature of the gravitational setup that relates them. And in particular, this feature that in AdS$_3$ the level and the central charge are related in this way will play an important role in the following discussion.

As we did for the other examples, it is instructive to discuss unitary representations of the algebra. A primary of \eqref{eq:sl2KMalgebra} is defined as a state $|m,h;j\rangle$ that is an eigenstate of the zero modes
\begin{equation}
J_0^0 |m,h;j\rangle = m |m,h;j\rangle~, \qquad L_0 |m,h;j\rangle = h|m,h;j\rangle~,
\end{equation}
in addition to the quadratic Casimir of $sl(2,\RR)$
\begin{equation}
\eta_{ab} J^{a}_0J^{b}_0|m,h;j\rangle = -j(j-1) |m,h;j\rangle,
\end{equation}
and is also annihilated by $L_n$ and $J^a_n$ with $n>0.$ Descendent states are now created by acting with $L_{-n}$ and $J^a_{-n}$ ($n>0)$. In addition, discrete representations of $sl(2,\RR)$ typically fall into two classes; these are $\mathcal D_j^{(+)},$ which is defined by demanding $J_0^- |m,h;j\rangle =0,$ and $\mathcal D_j^{(-)}$ which has $J_0^+ |m,h;j\rangle = 0.$\footnote{Note that, because of the non-compact nature of $sl(2,\RR),$ the representations $\mathcal D_j^{(\pm)}$ contain an infinite set of states generated by the zero modes of $J_0^+$ or $J_0^-.$ Since, in thermal $AdS_3,$ the $J_0^\pm$ correspond to global elements of the symmetry algebra these $sl(2,\RR)$ descendents will not be seen in the gravity analysis for the graviton determinant.} See \cite{Maldacena2001a,Dixon1989} for a more detailed discussion of these representations.

One would expect to find that the graviton one-loop determinant for the $sl(2,\RR)$ KM boundary conditions arranges itself into a product of a Virasoro and $sl(2,\RR)$ KM character. The descendent contributions to an $sl(2,\RR)$ KM character take the form
\begin{equation}\label{eq:sl2char}
\chi^{(+)}_{sl(2)}(q,\bar q) = \frac{1}{1-\bar q}\prod^\infty_{n=1}\frac{1}{(1-q^n\bar q)(1-q^n)(1-q^n \bar q^{-1})},
\end{equation}
for a representation of the type $\mathcal D_{j}^{(+)}.$ Comparing this to (\ref{eq:gravdetSL2a}), we indeed find the appropriate structure, modulo the first factor in (\ref{eq:sl2char}) which corresponds to the $J_0^+$ descendent contribution. We recall that the product in (\ref{eq:gravdetSL2a}) corresponds to the graviton determinant where the ghost spectrum is treated with Dirchlet boundary conditions (i.e. the ghost fluctuations are strictly subleading relative to the spin-2 modes).

There is, however, a problem with the above analysis, which can be seen most easily by considering the sign in front of the $sl(2,\RR)$ level $k$ in the $[J_n^0,J_m^0]$ commutator. Since $k = c/6$ is positive, representations of the current algebra \eqref{eq:sl2KMalgebra} necessarily contain negative norm states. This is however not a problem: there is a natural interpretation as to why $k$ must appear precisely as in \eqref{eq:sl2KMalgebra}. The boundary theory dual to $sl(2,\RR)$ KM boundary conditions is a theory of induced gravity \cite{Apolo2014}. In particular, this theory is described by a two-dimensional induced gravity in light-cone gauge as originally formulated in \cite{Polyakov1987,Knizhnik1988}. 

As discussed in \cite{Apolo2014}, the appropriate boundary stress tensor includes the twisted Sugawara term, which amounts to a shift of the form
\begin{equation}
\hat T_{++}(t^+) = T_{++}(t^+) + \partial_+ J^0(t^+).
\end{equation}
This introduces the following shift in the Virasoro generators\footnote{We have also included a zero mode shift of $-\frac{c}{24}$ which can be thought of as arising from mapping the Virasoro generators $L_n$ on the plane to those on the cylinder $\hat L_n.$}
\begin{equation}
\hat L_n = L_n - i n J^0_n - \frac{c}{24} \delta_{n,0}.
\end{equation}
In terms of $\hat L_n$ one can check that the Virasoro algebra becomes
\begin{equation}
[\hat L_n,\hat L_m] = (n-m) \hat L_{n+m},
\end{equation}
where the shift by $J^0_n$ has lead to a cancellation between the bare central charge in (\ref{eq:sl2KMalgebra}) and a central term induced by the $sl(2,\RR)$ level $k=c/6.$ Note that it was crucial that the sign in front of $k$ in (\ref{eq:sl2KMalgebra}) is as written, otherwise the induced central charge would not have canceled the bare central term.

Since the twisted generators satisfy a Virasoro algebra with vanishing central charge, we can gauge the diffeomorphisms on the boundary. Now, we can see that the extra ghost states were necessary in order to arrive at the result in (\ref{eq:gravdetSL2}). These extra ghost states correspond precisely to the boundary diffeomorphisms, which remove all of the Virasoro $sl(2,\RR)$ KM descendent contributions, as expected when the dual 2d theory is a theory of gravity. 

\section{Discussion}\label{sec:disc}

In this work we have computed the pole structure of the graviton one-loop determinant in three dimensional AdS gravity with the aim of quantifying how chiral boundary conditions affect the determinant. In the following we discuss some important features of our results and some possible future directions.

\subsection*{Extensions of the DHS method}

We extended in three directions the quasinormal mode method first developed in \cite{DenefHartnollSachdev2010}. 

The first extension is the treatment of stationary, as opposed to static, spacetimes, which is required in order to distinguish between holomorphic and anti-holomorphic contributions to the graviton determinant. In the static case, poles in the one-loop determinant arise when the quasinormal mode frequencies are tuned to be proportional to the Euclidean thermal mode number.  Our primary result here is (\ref{eq:omegarot}), which shows that in the rotational case the quasinormal mode frequencies must instead be tuned to a particular combination of the Euclidean thermal mode number and the angular frequency.  Although this particular expression is specific to the BTZ black holes we study, we expect that the derivation procedure will be similar for other stationary spacetimes in any number of dimensions.

Our second extension concerns an improved treatment of fields with spin in the quasinormal mode method.  Although fields with spin have been studied previously in e.g. \cite{DattaDavid2012,KeelerLisbaoNg2016}, in appendices \ref{app:spin2Eucl} and \ref{app:spin1Euc} we provide a comprehensive discussion of the adjusted integer ranges required in those prior works. We show that for fields with spin, not every quasinormal mode Wick-rotates to a normalizable Euclidean mode.  For quasinormal modes with quantum number at or below the field's spin, the thermal mode number may have a restricted range in order to achieve normalizability at the tip of the Euclidean cigar and thus a pole in the one-loop determinant.  And although we have studied a particular example, we expect that this subtlety will generalize to any scenario where DHS is applicable.

The most obvious extension required in our work is to apply the quasinormal mode method to the case of chiral boundary conditions. These impose Neumann boundary conditions on the left-moving graviton components, while the right-moving components remain Dirichlet, as detailed in section \ref{sec:bc}.  We applied the DHS procedure to this situation, and found reasonable results. For massive spin fields determining the modes that contribute to the determinant requires some work but is straightforward; this is done in section \ref{sec:ms2d} for the spin-2 field, and generalizations should follow naturally. The more interesting feature appears for massless fields and their ghost contribution which we discuss below.

\subsection*{Holography going wild at the boundary}
We focused on two types of chiral boundary conditions, distinguished by a particular functional constraint on the boundary metric. Allowing the left-moving components of the boundary metric to vary as an arbitrary function of the boundary coordinates, one finds the asymptotic symmetry algebra contains an $sl(2,\RR)$ current algebra \cite{Avery2014}. As argued in \cite{Apolo2014} the holographic dual of these boundary conditions corresponds to two-dimensional gravity in a chiral light-cone gauge as in \cite{Polyakov1987,Knizhnik1988}. Our results for the graviton determinant in \eqref{eq:gravdetSL2} confirm these expectations by demonstrating that the Virasoro and $sl(2,\RR)$ descendants are removed from the spectrum. 

The second type of boundary conditions we considered are the more stringent ones of CSS \cite{CompereSongStrominger2013a}. These conditions require the fluctuating boundary metric to depend only on the left-moving coordinate $t^+$; they produce a Virasoro $U(1)$ Kac-Moody asymptotic symmetry algebra. As proposed in \cite{CompereSongStrominger2013a} the holographic interpretation in this case is in terms of a warped conformal field theory and as described in Section \ref{sec:CSShol} our result in (\ref{eq:gravdetCSS1}) for the graviton determinant reinforces this idea.

There are two interesting directions to explore here. One direction is to complement our analysis with the recent work in \cite{McGoughMezeiVerlinde2016,GiveonItzhakiKutasov2017}. There a deformation of the action can be interpreted as a modification of the boundary conditions in AdS, which also provides an interesting holographic interpretation.  Another direction is to explore the behaviour of one-loop determinants for other boundary conditions in AdS$_3$ such as those discussed recently in \cite{Grumiller2016,PerezTempoTroncoso2016} and references within.

\subsection*{Ghosts are scary}

For massless fields, such as the graviton, gauge invariance requires the introduction of ghost fields in the path integral. The Neumann nature of the chiral boundary conditions brings a subtlety to the ghost determinant as detailed in section \ref{sec:mdngraviton}, which we summarize here.

What are the appropriate boundary conditions for the vector ghost? We can either allow ghosts whose metric variations are on par with the allowed graviton modes, or instead require them to be purely subleading. This crucial distinction arises because allowing Neumann conditions for the ghost eigenfunctions opens up the possibility of including ghost states that actually gauge away the physical boundary gravitons. 

The choice of ghost boundary conditions for the $sl(2,\RR)$ KM case is rather natural. Unitarity of the boundary theory requires the inclusion of ghost states which grow at the boundary in order to cancel negative norm descendent states arising from the non-compact $sl(2,\RR)$ current algebra. For CSS, the choice of ghost boundary conditions is more subtle. An interpretation in terms of a WCFT requires that we do not allow for ghost modes which grow at the boundary, as in our result in (\ref{eq:gravdetCSS1}). However, within our framework, it is apparently just as valid to include some of the Neumann ghost states which, as in the $sl(2,\RR)$ case, remove the descendent states from the spectrum as we found in (\ref{eq:gravdetCSSgauged}). A full understanding of the holographic interpretation of this case is still lacking, although the chiral Liouville gravity of \cite{CompereSongStrominger2013} will likely play a role. 

All of these cases illustrate the importance in defining the physical states corresponding to the boundary gravitons and identifying the appropriate conditions on the ghost eigenfunctions. For this purpose, it would be very useful to develop a gauge invariant procedure for constructing the one-loop determinant which does not require the introduction of ghost modes, but we leave this for future work.

\subsection*{The entire function ${\rm Pol(\Delta)}$}

 We have chosen to study only the pole structure of the one-loop determinant and thus have ignored the $e^{\rm Pol(\Delta)}$ factor in (\ref{eq:GMY}).  The purpose of the polynomial factor is to account for zero modes and renormalization effects, including the multiplicative anomaly \cite{Cognola:2014pha}. This choice to focus on the pole structure of the one-loop determinant alone does mean we cannot compute, e.g., the Casimir energy as noted in footnote \ref{qhbarqPfootnote}.

The choice of chiral boundary conditions complicates the calculation of the function $e^{\rm Pol(\Delta)}$. In the case of Dirichlet conditions, this factor can be found by comparing the large $\Delta$ behavior of the pole structure to, e.g., the large $\Delta$ behavior required by the heat kernel curvature expansion as in \cite{DenefHartnollSachdev2010}.  In the case of pure Neumann conditions, a similar result could be found by instead studying the $\Delta \rightarrow -\infty$ limit;  however, in the chiral conditions we consider, we have both Dirichlet and Neumann modes, so neither limit is easy to study.  It might be possible to divide the infinite product into definite helicity sectors similar to the approach taken in \cite{Arnold:2016dbb} which factored the determinant into fixed momentum sectors, but we leave any such consideration to future work.

\section*{Acknowledgements}

It is a pleasure to thank P.~Betzios, F.~Larsen, G.~S\'{a}rosi, W.~Song, and D.~Vaman for useful discussions. The work of A.C. and P.S. is part of the Delta ITP consortium, a program of the NWO that is funded by the Dutch Ministry of Education, Culture and Science (OCW); it was also performed in part at the Aspen Center for Physics, which is supported by National Science Foundation grant PHY-1607611. A.C. is supported by Nederlandse Organisatie voor Wetenschappelijk Onderzoek (NWO) via a Vidi grant. The work of C.K. is supported in part by the European Union's Horizon 2020 research and innovation programme under the Marie Sk\l{}odowska-Curie grant agreement No 656900, and in part by the Danish National Research Foundation project ``New horizons in particle and condensed matter physics from black holes". C.K. and P.S. would also like to thank the Centro de Ciencias de Benasque Pedro Pascual as well as the organizers of the workshop ``Gravity - New perspectives from strings and higher dimensions" for hospitality during the final stages of this project.


\appendix

\section{BTZ black hole in various coordinates}\label{app:btz}

In this appendix we compile several useful coordinate systems to describe the BTZ black hole; all equations have the AdS radius set to one.  We begin with the more traditional Boyer-Lindquist type coordinates
\be\label{eq:coordbl}
\frac{ds^2}{\ell^2} = {r^2 \over (r^2-r_+^2)(r^2-r_-^2)}dr^2- {(r^2-r_+^2)(r^2-r_-^2)\over r^2}dt^2 + r^2\le(d\phi-{r_+r_-\over r^2}dt\ri)^2~,
\ee
where as usual we have $\phi \sim \phi +2\pi$. The inner and outer horizons are related to the mass and angular momentum via
\be
r_+^2 +r_-^2 = M~, \qquad r_+^2 r_-^2 = \frac{J^2}{4}~.
\ee

In  Fefferman-Graham coordinates the BTZ takes the form
\be\label{CSScoord}
{ds^2} = {d\rho^2}-e^{2\rho} dt^+ dt^- + \frac{\left(r_+ + r_-\right)^2}{4}\left(dt^+\right)^2+ \frac{\left(r_+-r_-\right)^2}{4}\left(dt^-\right)^2-\frac{\left(r_+^2-r_-^2\right)^2}{16\ell^4}e^{-2\rho}dt^+ dt^-~,
\ee
where we have defined
\be
t^{\pm} = t \pm \phi~, \qquad r^2=r_+^2\cosh^2(\rho-\rho_0)-r_-^2\sinh^2(\rho-\rho_0)~, \qquad e^{2\rho_0}= {r_+^2 -r_-^2\over 4}~.
\ee

When performing the Euclidean continuation, it is most natural to make the following coordinate transformation:
\be\label{eq:defnTP}
\tanh^2 \xi = \frac{r^2-r_+^2}{r^2 - r_-^2}~, \qquad T = r_+t - r_-\phi~, \qquad \Phi = r_+ \phi - r_-t~.
\ee
In these coordinates, the metric is
\be
ds^2 = d\xi^2 - \sinh^2 \xi dT^2 + \cosh^2 \xi d\Phi^2~.
\ee
We refer to these as \emph{regular coordinates} because in terms of the Euclidean time coordinate, $T = -i T_E $, the metric becomes simply
\be \label{eq:regcoord}
ds^2 = d\xi^2 + \sinh^2 \xi dT_E^2 + \cosh^2 \xi d\Phi^2~,
\ee
and regularity at $\xi=0$ naturally fixes the periodicity of $T_E$ to be
\be
T_E \sim T_E + 2\pi~.
\ee
Note that the Euclidean continuation in the coordinates \eqref{eq:coordbl} implies that $r_-$ is purely imaginary and $t=-it_E$.
We will also find it occasionally useful to further transform the radial coordinate by 
\be
z= \tanh^2 \xi~,
\ee
in which case the metric is
\be
ds^2 = \frac{1}{4z(1-z)^2}dz^2 - \frac{z}{1-z}dT^2 + \frac{1}{1-z}d\Phi^2~.
\ee

Finally, when analyzing the massive spin-$1$ and spin-$2$ equations it is further useful to define dimensionless left-moving and right-moving coordinates
\bea\label{DDsec3.2}
&& x_L \equiv T + \Phi = (r_+-r_-)t^+~, \\ && x_R \equiv T - \Phi= (r_++r_-)t^-~,
\eea
in terms of which the various tensor components in the equations become diagonal. In these coordinates the metric is given by
\be
ds^2 = d\xi^2 - \fft12\cosh2\xi \, dx_L dx_R + \fft14(dx_L^2 + dx_R^2)~.
\ee

\section{Linearized graviton equations and quasinormal modes}\label{app:spin2}

In this appendix we will derive the massive spin-$2$ quasinormal spectrum for Dirichlet boundary conditions, which was originally done in \cite{DattaDavid2012}, and for chiral boundary conditions. We will also elaborate on the restrictions imposed on the Euclidean solutions, which affects the modes contributing to the determinants and are non-trivial for spin-$s$ fields (but trivial for scalar fields).

\subsection{Massive spin-$2$ equations}

As in equations \eqref{eq:firstorderspin2} and \eqref{eq:firstorderspin2both}, a massive spin-2 excitation $h_{\mu\nu}$ in AdS$_3$ satisfies the first order equation%
\footnote{Our notation differs from that used in  \cite{DattaDavid2012}.  In comparison to the coordinates used there, we have  $(x_{1},x_2)_{\rm there} = (x_{L},x_R)_{\rm here}$, and $(x^+,x^-)_{\rm there}=(T,\Phi)_{\rm here}$.} 
\be\label{eq:firstorderspin2APP}
\epsilon_\mu{}^{\alpha\beta}\nabla_{\alpha}h_{\beta\nu} = - m h_{\mu\nu}~,
\ee
which is equivalent to
\begin{align}
\nabla^\mu h_{\mu\nu}= 0~,\quad
h^\mu{}_\mu = 0 ~, \quad
\nabla^2 h_{\mu\nu} =  (m^2 - 3)h_{\mu\nu}~,
\end{align}
where we have set the AdS radius to one. To avoid cluttering, in this appendix we are dropping the subscript in $m$ (in the main text it is denoted as $m_2$). 

Using the tracelessness condition
\begin{equation}\label{eq:traceless}
h_{\xi\xi} = \frac{1}{\sinh^2\xi} h_{TT} -\frac{1}{\cosh^2\xi} h_{\Phi\Phi}~,
\end{equation}
and the first order equations of motion, one can solve algebraically for the components $h_{\xi\xi}, \, h_{\xi T},\, h_{\xi \Phi},$ and thus express the equations of motion solely in terms of the components of $h_{\mu\nu}$ along the boundary directions. It is at times useful to express the remaining spin-$2$ tensor components in the $(x_L,x_R)$ basis, whereas at other times it is convenient to express them in the $(T,\Phi)$ basis. We will use both often, and the relation between them reads
\begin{equation}\label{eq:hcomprelations}
\begin{pmatrix}
h_{TT} \\
h_{T\Phi} \\
h_{\Phi\Phi} 
\end{pmatrix}
=
\begin{pmatrix}
1 &  2 &  1  \\
1 &  0 & -1  \\
1 & -2 &  1
\end{pmatrix}
\begin{pmatrix}
h_{LL} \\
h_{LR} \\
h_{RR} 
\end{pmatrix}.
\end{equation}

We Fourier expand the spin-$2$ field as
\begin{equation}\label{eq:hansatz}
h_{\mu\nu}(z,T,\Phi) = e^{- i (k_L x_L + k_R x_R)} R_{\mu\nu}(z)~.
\end{equation}
Just as we use either the $(x_L,x_R)$ or $(T,\Phi)$ basis for the spin-$2$ components above, we will find it useful below to express the momentum with respect to the several different choices of coordinates. The relation between the various definitions follows from 
\begin{align}
e^{-i(\omega t - k \phi)} = e^{-i(k_L x_L + k_R x_R)} = e^{-i(k_T T + k_\Phi \Phi)}~,
\end{align}
which implies the relations
\begin{align}\label{eq:klkr}
\omega - k = 4\pi T_L k_L~, \qquad & \omega + k = 4\pi T_R k_R~, \\
k_T = k_L + k_R~, \qquad & k_\Phi = k_L - k_R~.
\end{align}
In addition, because $\phi$ parameterizes a circle, regularity of the solutions implies $k \in \mathbb{Z}$.
In the $(x_L,x_R)$ basis, the equations of motion for the radial wave functions become diagonal. In particular, one has \cite{DattaDavid2012}
\begin{eqnarray}
z(1-z) \frac{d^2 R_{LL}}{dz^2} + (1-z) \frac{d R_{LL}}{dz} + \left[\frac{k_T^2}{4z}-\frac{k_\Phi^2}{4} - \frac{(m+2)^2-1}{4(1-z)}\right]R_{LL} = 0~, &&\\
z(1-z) \frac{d^2 R_{LR}}{dz^2} + (1-z) \frac{d R_{LR}}{dz} + \left[\frac{k_T^2}{4z}-\frac{k_\Phi^2}{4} - \frac{m^2-1}{4(1-z)}\right]R_{LR} = 0~, &&\\
z(1-z) \frac{d^2 R_{RR}}{dz^2} + (1-z) \frac{d R_{RR}}{dz} + \left[\frac{k_T^2}{4z}-\frac{k_\Phi^2}{4} - \frac{(m-2)^2-1}{4(1-z)}\right]R_{RR} = 0~. &&
\end{eqnarray}
The solutions to these equations are given by
\begin{eqnarray}\label{eq:spin2gensol}
R_{ij}(z) &=& z^{-\frac{i}{2} k_T}R_{ij}^\tin(z) + z^{\frac{i}{2} k_T}R_{ij}^\tout(z) \\
&=& (1-z)^{\beta_{ij}} \left[ e^{\text{in}}_{ij} z^{-\frac{i}{2} k_T} F\left(a^{\text{in}}_{ij},b^{\text{in}}_{ij},c^{\text{in}};z\right) + e^{\text{out}}_{ij} z^{\frac{i}{2} k_T} F\left(a^{\text{out}}_{ij},b^{\text{out}}_{ij},c^{\text{out}};z\right)\right]~,
\end{eqnarray}
where we have written the solutions such that the the functions $R_{ij}^\tin(z)$ and $R_{ij}^\tout(z)$ become unity at the horizon $z=0.$ 
The sign of the exponent of $z$ indicates that the ``in" and ``out" superscripts naturally refer to ingoing and outgoing solutions. $e^{\text{in}}_{ij}$ and $e^{\text{out}}_{ij}$ are polarization constants and the other constant parameters are given by
\begin{eqnarray}
&&\beta_{LL} = \frac{m+3}{2},  \qquad \beta_{LR} = \frac{m+1}{2}~, \qquad \beta_{RR} = \frac{m-1}{2}~, \\
&& a^{\text{in}}_{ij} = -i k_R + \beta_{ij}, \qquad b^{\text{in}}_{ij} = -i k_L + \beta_{ij}~, \qquad c^{\text{in}} = 1 - i(k_L+k_R)~, \\
&& a^{\text{out}}_{ij} = i k_L + \beta_{ij}, \qquad b^{\text{out}}_{ij} = i k_R + \beta_{ij}~, \qquad c^{\text{out}} = 1 + i(k_L+k_R)~.
\end{eqnarray}
For ingoing solutions (with $e_{ij}^\tout = 0$), the polarization constants are constrained by the first-order equations to satisfy
\begin{eqnarray}\label{eq:eijrelationsin}
(m+1+2ik_R)e^\tin_{LL} &=& -(m+1-2ik_L)e^\tin_{LR}~,\\
(m-1+2ik_R)e^\tin_{LR} &=& -(m-1-2ik_L)e^\tin_{RR}~,
\end{eqnarray}
whereas for the outgoing solutions (with $e_{ij}^\tin = 0$) one has
\begin{eqnarray}\label{eq:eijrelationsout}
(m+1-2ik_R)e^\tout_{LL} &=& -(m+1+2ik_L)e^\tout_{LR}~,\\
(m-1-2ik_R)e^\tout_{LR} &=& -(m-1+2ik_L)e^\tout_{RR}~.
\end{eqnarray}
Note that the ingoing (radial) wave-function is simply related to the corresponding outgoing one by sending $(k_L,k_R) \rightarrow -(k_R,k_L).$ 

\subsection{Determining the spectra}

From now on we focus on the ingoing solutions. Writing them out explicitly, we have
\begin{eqnarray}\label{eq:RinSol}
R_{LL}(z) &=& e^{\text{in}}_{LL} (1-z)^{\frac{m+3}{2}}z^{-\frac{i}{2} k_T}  F\left(a^{\text{in}}_{LL},b^{\text{in}}_{LL},c^{\text{in}};z\right)~  ,\\
R_{LR}(z) &=& e^{\text{in}}_{LR}(1-z)^{\frac{m+1}{2}}  z^{-\frac{i}{2} k_T} F\left(a^{\text{in}}_{LR},b^{\text{in}}_{LR},c^{\text{in}};z\right)~ , \\
R_{RR}(z) &=& e^{\text{in}}_{LR}(1-z)^{\frac{m-1}{2}} z^{-\frac{i}{2} k_T} F\left(a^{\text{in}}_{RR},b^{\text{in}}_{RR},c^{\text{in}};z\right)~.
\end{eqnarray}
In order to relate the ingoing wave-function to an expansion at the boundary we use the connection identity 
\begin{eqnarray}\label{eq:HGz=1}
F(a,b;c;z) &=& \frac{\Gamma(c)\Gamma(c-a-b)}{\Gamma(c-a)\Gamma(c-b)} F(a,b;a+b-c+1;1-z) \nn\\
&&+ (1-z)^{c-a-b} \frac{\Gamma(c)\Gamma(a+b-c)}{\Gamma(a)\Gamma(b)}F(c-a,c-b;c-a-b+1;1-z)~.
\end{eqnarray}
Near the boundary, $z \to 1$ and $r^{-2}\sim 1-z \to 0$. Using the connection formula (\ref{eq:HGz=1}) to expand the solutions (\ref{eq:RinSol}) for large $r,$ assuming $m>0$ we find the following behavior:
\begin{eqnarray}\label{eq:Rmplusin}
R_{LL} &\simeq& e^\tin_{LL}r^{m+1} \frac{\Gamma(c^\tin)\Gamma(m+2)}{\Gamma(a^\tin_{LL})\Gamma(b^\tin_{LL})} (1+ \cdots) + \mathcal O\left(r^{-m-3}\right)~, \\
R_{LR} &\simeq& e^\tin_{LR}r^{m-1}  \frac{\Gamma(c^\tin)\Gamma(m)}{\Gamma(a^\tin_{LR})\Gamma(b^\tin_{LR})}\left(1 + \cdots\right) + \mathcal O\left(r^{-m-1}\right)~, \\
R_{RR} &\simeq& e^\tin_{RR}r^{-m+1} \frac{\Gamma(c^\tin)\Gamma(2-m)}{\Gamma(c^\tin-a^\tin_{RR})\Gamma(c^\tin-b^\tin_{RR})} \left(1 + \cdots\right) + \mathcal O\left(r^{m-3}\right)~. 
\end{eqnarray}
Notice that the expansion of $R_{RR}$ appears different in structure from the other components. This is because we have assumed that\footnote{The condition $|m|\geq1$ corresponds to the unitarity bound $\Delta \geq 2$ for spin-$2$ operators in the dual CFT.}
\begin{equation}
1 \leq |m| < 2~,
\end{equation}
which contains the value $m = 1,$ corresponding to the graviton. For positive values of $m$ with $|m|>2,$ the two series in the expansion of $R_{RR}$ swap dominance. When $m$ is negative, a similar statement applies. In particular, for $m < 0,$ the relevant expansion is
\begin{eqnarray}\label{eq:Rmminusin}
R_{LL} &=& e^\tin_{LL} r^{m+1}  \frac{\Gamma(c^\tin)\Gamma(m+2)}{\Gamma(a^\tin_{LL})\Gamma(b^\tin_{LL})}\,(1+\cdots) + \mathcal{O}(r^{-m-3})~, \\
R_{LR} &=& e^\tin_{LR}\,r^{-m-1} \frac{\Gamma(c^\tin)\Gamma(-m)}{\Gamma(c^\tin-a^\tin_{LR})\Gamma(c^\tin-b^\tin_{LR})}\left(1+ \cdots\right) + \mathcal{O}(r^{m-1}) ~,\\
R_{RR} &=& e^\tin_{RR}\,r^{-m+1} \frac{\Gamma(c^\tin)\Gamma(2-m)}{\Gamma(c^\tin-a^\tin_{RR})\Gamma(c^\tin-b^\tin_{RR})}(1+\cdots) + \mathcal{O}(r^{m-3})~,\label{eq:RRRmminusin}
\end{eqnarray}
in which case we see that the two series in the expansion of $R_{LL}$ swap dominance for $m<-2.$

\subsubsection{Quasinormal Boundary Conditions}

Assuming the condition
\begin{equation}
1 \leq |m| < 2~,
\end{equation}
the standard quasinormal boundary conditions correspond to enforcing that the leading divergence in the boundary expansions in (\ref{eq:Rmplusin}) or (\ref{eq:Rmminusin}) vanish; this ensures that the perturbation is normalizable as $r\rightarrow \infty.$  For $m>0$ $(m<0),$ this corresponds to demanding the leading term in $R_{LL}$ $(R_{RR})$ vanish. 

For $m>0$ we find the ingoing quasinormal spectrum to be
\begin{equation}\label{eq:QNMmplusin}
\begin{drcases}
\, 2ik_R = 2p + \Delta + 2  \,\,\, \\
\,2ik_L =  2p + \Delta - 2  \,\,\,
\end{drcases}\,\,\,
\text{for all integers} \,\,\, p\ge 0~,
\end{equation}
where we have defined $\Delta = |m| + 1.$ We will refer to $p$ as the radial quantum number. Almost all of these modes arise by ensuring the $\Gamma$-functions in the denominator of (\ref{eq:Rmplusin}) acquire poles which set the leading term in $R_{LL}$ to zero. This vanishing occurs when either $a_{LL}^\tin$ or $b_{LL}^{\tin}$ becomes equal to zero or a negative integer. However, there are two special solutions, corresponding to $p=0,\,1$ in the $k_L$ series. These solutions instead have parameters set such that the polarization tensor component $e^\tin_{LL}$ in (\ref{eq:eijrelationsin}) vanishes. There are two possibilities, corresponding to setting $p=0$ and $p=1$ in the second line of (\ref{eq:QNMmplusin}).

For $m<0,$ there is a similar story which imposes conditions on the leading behavior of $R_{RR}$ in (\ref{eq:RRRmminusin}). We find the modes
\begin{equation}\label{eq:QNMmminusin}
\begin{drcases}
\, 2ik_R = 2p + \Delta - 2  \,\,\, \\
\, 2ik_L =  2p + \Delta + 2  \,\,\,
\end{drcases}\,\,\,
\text{for all integers} \,\,\, p\ge 0~.
\end{equation}
The outgoing solutions can also be handled similarly. In the end we arrive at the quasinormal mode spectrum displayed in Table \ref{table:BHQNMspin2}. 

\subsubsection{Chiral boundary conditions}\label{app:mixedQNM}

We will now consider the chiral  boundary conditions relevant for the analysis in section \ref{sec:mixed}. In particular, for the graviton with $|m|=1,$ one imposes that $R_{RR}$ falls off faster than $\mathcal{O}(r^0)$ at the boundary while allowing $R_{LL}$ to fluctuate at $\mathcal{O}(r^2).$ The boundary condition thus amounts to demanding conditions solely on $R_{RR}.$ These conditions have a natural continuation for $m$ in the range 
\begin{equation}
1 \leq |m| < 2~.
\end{equation}

Let us examine the behavior of the wave-functions in (\ref{eq:Rmplusin}) and (\ref{eq:Rmminusin}). For $m<0,$ requiring the leading term in $R_{RR}$ to vanish is the same condition we required in the previous subsection, so the chiral boundary conditions for $m<0$ are implemented in the same way as the standard quasinormal condition. However, for $m>0$ the chiral boundary conditions place restrictions on the asymptotic behavior of $R_{RR}$ instead of $R_{LL}$ as was the case for the standard boundary conditions. This means that for $m>0$ we require $c_{RR}^\tin - a^\tin_{RR}$ or $c_{RR}^\tin - b^\tin_{RR}$ to be zero or a negative integer. In addition, there are again two special conditions arising from setting $e_{RR}^\tin =0$ in (\ref{eq:eijrelationsin}). The entire spectrum satisfying chiral  boundary conditions is presented in Table \ref{table:mixedQNMspin2}.

\subsection{Regularity of Euclidean solutions}\label{app:spin2Eucl}

The mode functions defined in (\ref{eq:hansatz}) have a natural continuation to Euclidean signature. In this section we use the coordinates (\ref{eq:regcoord}), where regularity at the origin $\xi=0$ is made most manifest. At the level of the solutions to the wave equation, the Euclidean continuation is implemented by making the replacements
\begin{equation}
T = - i T_E~, \qquad k_T = i k_E~,
\end{equation}
and the periodicity in $T_E$ constrains the values of $k_E$ such that
\begin{equation}
k_E \in \mathbb Z~.
\end{equation}
Setting $k_T = ik_E$ in the solutions (\ref{eq:spin2gensol}), we see that normalizability at small\footnote{Note that small $\xi$ corresponds to small $z$, where $z\sim \xi^2.$} $\xi$ naturally identifies positive values of $k_E$ with the ingoing solutions, such that one sets
\begin{equation}
k_T = i k_E = i n~, \qquad n > 0~.
\end{equation}
Correspondingly, the negative values of $k_E$ are assigned to the outgoing solutions, with
\begin{equation}
k_T = i k_E = i n~, \qquad n < 0~.
\end{equation}
In addition, one can consider the zero modes 
\begin{equation}
k_T = i k_E = 0~,
\end{equation}
as arising from either sector.

Finally, for some specific states, there is an additional restriction on the allowed values of $n.$ This restriction arises from demanding square-integrability of the Euclidean solutions near $\xi=0$. In particular, we demand that the Euclidean solutions $h^{(\lambda)}_{\mu\nu}$ satisfy \cite{heatkernel4}
\begin{equation}\label{eq:sqint}
\int d^3 x\sqrt{g} g^{\mu\nu}g^{\rho\sigma}h^{(\lambda)}_{\mu\rho}(x) h^{(\lambda')*}_{\nu\sigma}(x) = \delta(\lambda-\lambda')~,
\end{equation}
where $\lambda$ is an eigenvalue and the asterisk denotes complex conjugation. In order to avoid a non-integrable singularity at $\xi = 0$ in the integrand of (\ref{eq:sqint}), we must further restrict the range of $n$ for Euclidean solutions with certain low-lying values of the radial quantum number $p.$ 

One can see that a potential problem exists by considering the component $h_{\xi\xi}$ which, since the inverse metric component $g^{\xi\xi} = 1,$ shows up squared with only the metric determinant as prefactor in (\ref{eq:sqint}). The tracelessness condition (\ref{eq:traceless}) implies that near the origin 
\begin{eqnarray}
h_{\xi\xi} &\sim& \frac{1}{\xi^2} h_{EE} \nn\\
&\sim& e_{EE} \xi^{|n| - 2}( 1 + \mathcal{O}(\xi))~,
\end{eqnarray}
where $h_{EE}$ and $e_{EE}$ are the Euclidean rotation of $h_{TT}$ and $e_{TT},$ which are related to the $(L,R)$ basis by the matrix equation (\ref{eq:hcomprelations}). This means that, for small $\xi,$ one has
\begin{equation}
\sqrt{g} g^{\mu\nu}g^{\rho\sigma}h^{(\lambda)}_{\mu\rho}(x) h^{(\lambda')*}_{\nu\sigma}(x) \sim e^2_{EE} \xi^{2k_E - 3}~.
\end{equation}
Therefore, for $k_E = 0,\,1$ there is a potential non-integrable singularity at $\xi=0.$

The potential singularity at $\xi=0$ is avoided for most values of $p$ because $e_{EE}$ vanishes for $k_E=0$ or $k_E=1$ in generic solutions.\footnote{One can check this by noticing that the polarization tensors satisfy the same matrix equation as the tensor components in (\ref{eq:hcomprelations}), where $h_{TT}$ and $h_{T\Phi}$ are related to $h_{EE}$ and $h_{E\Phi}$ by analytic continuation.  Imposing the relations (\ref{eq:eijrelationsin}) or (\ref{eq:eijrelationsout}), one sees that indeed $e_{EE}$ vanishes where $k_E=0,$ $k_E= 1$ (for ingoing) or $k_E=-1$ (for outgoing).} However, there are a finite number of states where this is not satisfied. In particular, focusing on the $m>0$ states,  we find that the Euclidean continuations of states belonging to the $k_L$ series in Table \ref{table:BHQNMspin2} and to the $k_R$ series in Table \ref{table:mixedQNMspin2} with mode numbers given by 
\begin{eqnarray}\label{eq:divsols}
(p,k_E) \in \{(0,0), \, (0,1),\, (0,-1), \, (1,0)\}~,
\end{eqnarray}
do not satisfy $e_{EE}=0.$ These modes correspond to wave-functions that are not square-integrable and should be discarded. We could also argue these states should be eliminated because they correspond to the special values of $p$ where components of the polarization tensors $e_{ij}$ vanish, as described following (\ref{eq:QNMmplusin}), for which $e_{EE} \neq 0.$ 

The states with quantum numbers (\ref{eq:divsols}) should also be discarded from the Euclidean continuation of the $k_R$ series with $m<0$ in Table \ref{table:BHQNMspin2} and Table \ref{table:mixedQNMspin2}. By shifting $p$ for the specific case of $n=- 1$ in this series, we can combine the $m<0, n<0$  $k_R$ series with the $m<0, n\geq0$ $k_L$ series; this combination results in the bottom row of the $m<0$ column in Table \ref{table:BHEuclCond}, now valid for all integers $n,k$ and $p\geq 0$.  By similarly shifting $p$ to exclude the rest of the singular solutions, we are left with the entire set of possible conditions on allowed Euclidean solutions which are presented for standard Brown-Henneaux boundary conditions in Table \ref{table:BHEuclCond} and for the chiral boundary conditions in Table \ref{table:mixedEuclCond}.

\subsection{Checking the $\delta g_{r+}$ behavior}\label{app:gr+}

Finally, we need to understand the consequences of the $\delta g_{r+}$ condition in the two sets of boundary conditions in (\ref{eq:allowed3d}) and (\ref{eq:allowed3dSL2}). We begin by solving the first order equations (\ref{eq:firstorderspin2APP}), finding\footnote{Recall from (\ref{DDsec3.2}) that $x_L = (r_+-r_-)t^+.$}
\begin{equation}
h_{\xi L} = \frac{i}{k_R+k_L\cosh 2\xi}\left(\cosh2\xi\partial_\xi h_{LL} + \partial_\xi h_{LR} - (m+1)\sinh2\xi h_{LL} \right)~.
\end{equation}
Inserting the generic solutions (\ref{eq:spin2gensol}) into this expression and using that near the boundary $e^\xi \sim r$ and $h_{\xi L} \sim r h_{r+}$ we find that for large $r$ and $m>0$
\begin{equation}
h_{r+} \sim r^{m-2} e^\tin_{LL}\frac{k_R }{\Gamma(a^\tin_{LL})\Gamma(b^\tin_{LL})}\left(1 + \cdots \right) + O(r^{-m-3})~.
\end{equation}
For $m=1,$ this means that the leading behavior of $h_{r+} \sim O(r^{-1}),$ which is consistent with the $sl(2,\RR)$ Kac-Moody boundary conditions in (\ref{eq:allowed3dSL2}). However, CSS boundary conditions require $h_{r+} \sim O(r^{-3})$ for $m=1,$  and this condition is met only when either one of
\begin{equation}\label{eq:hr+cond1}
a^\tin_{LL} = -p~, \qquad b^\tin_{LL} = -p~, \qquad e^\tin_{LL}=0~,
\end{equation} 
is satisfied, or
\begin{equation}\label{eq:hr+cond2}
k_R = 0~.
\end{equation}
Notice that (\ref{eq:hr+cond1}) are precisely the Brown-Henneaux conditions. This means that the CSS boundary conditions can only be consistent if, in addition to the Neumann conditions described in section \ref{app:mixedQNM}, either the Brown-Henneaux conditions are satisfied or $k_R=0.$ 

\section{Analysis of ghost contributions to the gravitational path integral}\label{app:spin1}

In this appendix we present a detailed analysis of the a massive spin-1 field in AdS$_3$, and the ghost determinant that appears in the graviton one-loop path integral.

\subsection{Spin-1 equations}

The massive spin-$1$ modes can be solved similarly to the massive spin-$2$ modes. In first order form, the equation of motion is 
\begin{equation}\label{eq:firstorderspin1}
\epsilon_\mu{}^{\nu\rho} \nabla_\nu V_\rho = - m V_\mu~.
\end{equation}
Again we drop indices on $m$ to avoid clutter (in the main text it would be $m_1$). The solutions of this equation satisfy the massive vector equations of motion
\begin{eqnarray}
(\nabla^\nu\nabla_\nu - m^2 + 2)V_\mu &=& 0~, \nn\\
\nabla^\mu V_\mu &=& 0~. \label{eq:ghosteom}
\end{eqnarray}
The specific value of $m$ which corresponds to the spin-$2$ ghost is then $m^2= 4,$ i.e. $m = \pm 2$. For this value of the mass, we interpret $V_\mu$ as variation of the metric:
\begin{equation}\label{eq:gvar}
\delta g_{\mu\nu} = \nabla_{\mu} V_\nu + \nabla_\nu V_\mu~.
\end{equation}

It will  be useful to switch between the various coordinates: the vector components in the $(z,x_L,x_R)$ coordinates are related to those in the $(\xi,T,\Phi)$ coordinates by
\begin{eqnarray}
V_L &=& \frac{1}{2}(V_T + V_\Phi)~, \nn\\
V_R &=& \frac{1}{2}(V_T -V_\Phi)~, \nn\\
V_z &=& \frac{1}{2} \cosh^2\xi \coth \xi \,V_\xi~.
\end{eqnarray}
In components, equation \eqref{eq:firstorderspin1} reads
\begin{eqnarray}\label{eq:FirstOrderVecEq}
-m V_\xi &=& \frac{i}{\sinh\xi\, \cosh\xi}(k_T V_\Phi - k_\Phi V_T)~, \\
-m V_T &=& -\tanh\xi (\partial_\xi V_\Phi + i k_\Phi V_\xi)~, \\
-m V_\Phi &=& -\coth\xi (\partial_\xi V_T + i k_T V_\xi)~.
\end{eqnarray}
We will again look for solutions of the form
\begin{equation}
V_\mu(z,T,\Phi) = e^{-i (k_L x_L + k_R x_R)}R_{\mu}(z)~.
\end{equation}
Equation \eqref{eq:FirstOrderVecEq} can be thought of as a constraint on $R_\xi,$ and the remaining equations imply
\begin{eqnarray}
z(1-z) \frac{d^2 R_{L}}{dz^2} + (1-z) \frac{d R_{L}}{dz} + \left[\frac{k_T^2}{4z}-\frac{k_\Phi^2}{4} - \frac{(m+1)^2-1}{4(1-z)}\right]R_{L} = 0~, &&\\
z(1-z) \frac{d^2 R_{R}}{dz^2} + (1-z) \frac{d R_{R}}{dz} + \left[\frac{k_T^2}{4z}-\frac{k_\Phi^2}{4} - \frac{(m-1)^2-1}{4(1-z)}\right]R_{R} = 0~. &&
\end{eqnarray}
These have solutions given by
\begin{eqnarray}
R_{L}(z) &=& (1-z)^{\frac{m+2}{2}} \left[ e^{\text{in}}_{L} z^{-\frac{i}{2} k_T} F\left(a^{\text{in}}_{L},b^{\text{in}}_{L},c^{\text{in}};z\right) + e^{\text{out}}_{L} z^{\frac{i}{2} k_T} F\left(a^{\text{out}}_{L},b^{\text{out}}_{1},c^{\text{out}};z\right)\right] ~,\\
R_{R}(z) &=& (1-z)^{\frac{m}{2}} \left[ e^{\text{in}}_{R} z^{-\frac{i}{2} k_T} F\left(a^{\text{in}}_{R},b^{\text{in}}_{R},c^{\text{in}};z\right) + e^{\text{out}}_{R} z^{\frac{i}{2} k_T} F\left(a^{\text{out}}_{R},b^{\text{out}}_{R},c^{\text{out}};z\right)\right]~,
\end{eqnarray}
where
\begin{eqnarray}
&&\beta_{L} = \frac{m+2}{2}~,  \qquad \beta_{R} = \frac{m}{2}~,  \\
&& a^{\text{in}}_{i} = -i k_R + \beta_{i}~, \qquad b^{\text{in}}_{i} = -i k_L + \beta_{i}~, \qquad c^{\text{in}} = 1 - ik_T~, \\
&& a^{\text{out}}_{i} = i k_L + \beta_{i}~, \qquad b^{\text{out}}_{i} = i k_R + \beta_{i}~, \qquad c^{\text{out}} = 1 + i k_T~.
\end{eqnarray}

These solutions are not independent; the first-order equations imply constraints between the polarization vector components $e_L$ and $e_R.$ The relations are different for ingoing and outgoing solutions and are given by
\begin{eqnarray}\label{eq:spin1polvecs}
\left[2ik_R + m\right]e^\tin_{L} &=& \left[2ik_L - m\right]e^\tin_{R}~, \\ 
\left[2ik_R - m\right] e^\tout_L &=& \left[2ik_L + m\right]e^\tout_R~.
\end{eqnarray}

Again, utilizing the $z \simeq 1$ expansion of the hypergeometrics in (\ref{eq:HGz=1}) we find the boundary behavior of the ingoing solutions to be
\begin{eqnarray}\label{eq:vectorbdryingoing}
R^\tin_{L} &=& e^\tin_{L} \bigg[(1-z)^{\frac{1}{2}(m+2)}\frac{\Gamma(c^\tin)\Gamma(-m-1)}{\Gamma(c^\tin-a^\tin_{L})\Gamma(c^\tin-b^\tin_{L})}F(a^\tin_{L},b^\tin_{L};m+2;1-z) \nn\\ 
&&  + (1-z)^{-\frac{m}{2}}\frac{\Gamma(c^\tin)\Gamma(m+1)}{\Gamma(a^\tin_{L})\Gamma(b^\tin_{L})}F(c^\tin-a^\tin_{L},c^\tin-b^\tin_{L};-m;1-z)\bigg]~, \\
R^\tin_{R} &=& e^\tin_{R} \bigg[(1-z)^{\frac{m}{2}}\frac{\Gamma(c^\tin)\Gamma(-m+1)}{\Gamma(c^\tin-a^\tin_{R})\Gamma(c^\tin-b^\tin_{R})}F(a^\tin_{R},b^\tin_{R};m;1-z) \nn\\ 
&&  + (1-z)^{-\frac{1}{2}(m-2)}\frac{\Gamma(c^\tin)\Gamma(m)}{\Gamma(a^\tin_{R})\Gamma(b^\tin_{R})}F(c^\tin-a^\tin_{R},c^\tin-b^\tin_{R};-m+2;1-z)\bigg]~.
\end{eqnarray}
Computing the induced metric perturbations in (\ref{eq:gvar}), we find that Dirichlet boundary conditions require
\begin{eqnarray}\label{eq:BHvector1}
a_L^\tin = -i k_R + \frac{m+2}{2} &=& - p~, \nn \\
b_L^\tin = -i k_L + \frac{m+2}{2} &=& - p + 1~,
\end{eqnarray}
where $p$ is a non-negative integer and the shift by one in the second line arises for the mode where we demand $e_L^\tin=0$ in \eqref{eq:spin1polvecs}.

In order to hold $\delta g_{--}$ fixed, chiral boundary conditions require one of the constraints
\begin{eqnarray}\label{eq:mixedvector1}
a_R^\tin = -i k_R + \frac{m}{2} &=& - p~, \nn \\
b_R^\tin = -i k_L + \frac{m}{2} &=& - p~,
\end{eqnarray}
for all integers $p\ge 0.$ Note that these conditions contain the Brown-Henneaux ghost contributions in \eqref{eq:BHvector1} as a subset. In fact, the only new state in (\ref{eq:mixedvector1}) is the $p=0$ state in the $a_R^\tin$ tower (there is also a corresponding new outgoing state in the $b_R^\tout$ tower). There is again an additional state that comes about by requiring the polarization constant $e_R$ vanishes altogether, which completely kills the component $V_R.$ For ingoing states, this demands
\begin{equation}
-2ik_R - m = 0 \qquad (\text{ingoing})~,
\end{equation}
while for outgoing we have
\begin{equation}
2ik_R - m = 0 \qquad (\text{outgoing})~.
\end{equation}
Finally, there are two more states not included in the above analysis. In particular, when 
\begin{equation}
k_R = 0~,
\end{equation}
the induced metric variation $\delta g_{RR}$ vanishes; see equations (\ref{eq:metricghostvars}) through \eqref{eq:metricghostvars2}. This occurs for both $m_1>0$ and $m_1<0$ and both of these states should be included in the analysis. 

The final results for the set of ghost states consistent with chiral boundary conditions on the metric are given in Table \ref{table:BHQNMspin1} and Table \ref{table:Neumannspin1modes}.

\subsection{The Euclidean Solutions}\label{app:spin1Euc}

\begin{table}
	\begin{center}
		\begin{tabular}{|c|c|}
			\hline
			$m>0$ & $m<0$ \\
			\hline
			$\begin{matrix}\, 2p + \Delta + |n+1| + ik_\Phi(n,k) = 0  \,\,\, \\
			\,  2p + \Delta + |n-1| - ik_\Phi(n,k) = 0   \,\,\,\end{matrix}$&  
			$\begin{matrix}\, 2p + \Delta + |n-1| + ik_\Phi(n,k) = 0   \,\,\, \\
			\,2p + \Delta + |n+1| - ik_\Phi(n,k) = 0 \,\,\,\end{matrix}$\\ 
			\hline 
		\end{tabular}\captionsetup{justification=raggedright}\caption{\small Conditions on the quantum numbers of spin-1 states with Brown-Henneaux boundary conditions. For $m>0$ this is a subset of the states which are consistent with the chiral boundary conditions.}\label{table:BHspin1modes}
	\end{center}
\end{table}

All that remains now is to understand the Euclidean solutions into which the spin-$1$ states Wick-rotate. The Euclidean rotation on the momentum is again given by
\begin{equation}
k_T = i k_E = i n~,
\end{equation}
where ingoing solutions require that $n>0$ for regularity, outgoing solutions require that $n<0,$ and the zero modes can again be obtained from either ingoing or outgoing conditions with $n=0.$ For the Brown-Henneaux states in Table \ref{table:BHQNMspin1}, the process is almost identical to the spin-$2$ discussion and we compile the conditions in Table \ref{table:BHspin1modes}.

We now analyze the new states given in Table \ref{table:Neumannspin1modes} that are consistent with the chiral boundary conditions. First, consider the states in the first row of Table \ref{table:Neumannspin1modes}. After the Euclidean rotation, we can write the set of states as
\begin{eqnarray}\label{eq:newmodesReg}
k_E + i k_\Phi + \Delta - 1 = 0, && \qquad k_E \geq 0~, \\
-k_E - i k_\Phi + \Delta - 1 = 0, && \qquad k_E < 0~. 
\end{eqnarray}
We need to check that all of these states are regular at the origin. In particular, we require that $V_\xi$ is smooth as $\xi\rightarrow 0$; from \eqref{eq:FirstOrderVecEq}  we have
\begin{equation}
V_\xi = -\frac{i}{\sinh\xi\cosh\xi}\left[(ik_E - k_\Phi)V_L - (ik_E + k_\Phi)V_R\right]~.
\end{equation}
For $|k_E|>0 $ regularity at the origin is guaranteed, because $V_L \sim V_R \sim \xi^{|k_E|}$ for small $\xi.$ For $k_E=0$ we must check more carefully. Near the origin, we can expand
\begin{equation}
V_{L,R} = e_{L,R} + \mathcal O(\xi)~,
\end{equation}
Plugging the relations satisfied by the modes in (\ref{eq:newmodesReg}) into the polarization constant relations, and evaluating at $k_E = 0,$ we find $e_L = - e_R$
for both ingoing and outgoing modes. Hence, near the origin and taking $k_E=0$, we have $V_\xi\sim \mathcal O(1)$. 

The contributions from the second and third rows of Table \ref{table:Neumannspin1modes} are more subtle. The Wick rotation of the modes in the second row gives:
\begin{eqnarray}\label{eq:New2}
k_E + i k_\Phi - (\Delta -1) = 0 && \qquad k_E \geq 0~,\\  
-k_E - ik_\Phi - (\Delta - 1) = 0 && \qquad k_E \leq 0~.\label{eq:New22}
\end{eqnarray}
Only a subset of these modes correspond to admissible ghost states. To see this we need to evaluate the induced gauge transformation of the metric from each ghost state. In particular, the pure-gauge metric perturbations can be written in terms of a solution to the spin-$1$ equations with $m=\pm2.$ The induced gauge transformations are given by
\begin{eqnarray}\label{eq:metricghostvars}
\delta g_{LL} &=& - 2ik_L V_L~, \\
\delta g_{LR} &=& -ik_L V_R - i k_R V_L - \sinh \xi\cosh\xi \, V_\xi~,\\
\delta g_{RR} &=& - 2ik_R V_R~,\\
\delta g_{L\xi} &=& \partial_\xi V_L - i k_L V_\xi - 2\coth2\xi\, V_L- 2\csch 2\xi \, V_R~ ,\\
\delta g_{R\xi} &=& \partial_\xi V_R - ik_R V_\xi - 2\coth2\xi\, V_R- 2\csch 2\xi \, V_L~,\\
\delta g_{\xi\xi} &=& -2 \partial_\xi V_\xi~.\label{eq:metricghostvars2}
\end{eqnarray}

The states in the second row of Table \ref{table:Neumannspin1modes} have $e_R=0,$ which means that the condition $e_R=-e_L$ cannot be satisfied (except for the trivial solution $e_R=e_L=0$) and so the $k_E=0$ states are obviously not regular. Second, the $|k_E|=1$ contribution needs to be analyzed carefully. The wave-functions for these states are particularly simple; for generic $k_E,$ we have 
\begin{eqnarray}\label{eq:Sol2}
V_L &=& (\cosh \xi)^{\Delta-1} (\tanh\xi)^{|k_E|}e^{k_Ex_E -i k_\Phi \Phi}~ ,\\
V_R &=& 0~, \\
V_\xi &=& \frac{1}{(\Delta-1)\sinh\xi\,\cosh\xi}\left[(k_E+ik_\Phi)V_L -(k_E-ik_\Phi)V_R  \right]\nn\\
&=& (\cosh\xi)^{\Delta-3}(\tanh\xi)^{|k_E|-1}e^{k_Ex_E -i k_\Phi \Phi} ~.
\end{eqnarray}
Since $V_R=0$ for these states it is fairly straightforward to write out the induced metric variations. From (\ref{eq:metricghostvars}) through \eqref{eq:metricghostvars2} we have
\begin{eqnarray}\label{eq:specmetricghostvars}
\delta g_{LL} &=& (k_E- ik_\Phi) V_L~,\\
\delta g_{LR} &=&  \ft{1}{2}(k_E + ik_\Phi) V_L - \ft{1}{\Delta-1} (k_E+i k_\Phi)V_L~,\\
\delta g_{RR} &=& 0~,\\
\delta g_{L\xi} &=& \partial_\xi V_L + \ft{1}{2}(k_E-ik_\Phi) V_\xi - 2\coth2\xi\, V_L ~,\\
\delta g_{R\xi} &=& \ft{1}{2}(k_E+ik_\Phi) V_\xi - 2\csch 2\xi \, V_L~,\\
\delta g_{\xi\xi} &=& 2 \partial_\xi V_\xi~.\label{eq:specmetricghostvars2}
\end{eqnarray}
Evaluating these on (\ref{eq:Sol2}) and using (\ref{eq:New2}), we find that all of the induced metric variations vanish when we set $\Delta = 3$ and $k_E=1.$ Therefore, the solutions satisfying (\ref{eq:New2}) and (\ref{eq:New22}) at $k_E =\pm1$ and $\Delta=3$ correspond (locally) to Killing vectors of the BTZ background. These are modes with zero eigenvalue of the ghost Laplacian in \eqref{eq:ghosteom} and should be excluded from the pole contribution to the determinant. The same phenomena also occurs for the $k_R=0$ states with $|k_E|=1$ and $m_1 = -2,$ in which case $V_L=0$ because the polarization vectors $e_L$ in \eqref{eq:spin1polvecs} vanish. One can check that these also give trivial metric variations. The appearance of Killing vectors in the ghost determinant of massless gauge fields in the bulk is generic when performing alternative quantization \cite{Giombi2013}. As explained in \cite{Giombi2013}, since these are zero modes, they must be treated separately and generate $N^{-n_0/2}$ contributions to the partition functions as opposed to poles.

\bibliographystyle{JHEP}
\bibliography{ref3d}

\end{document}